\documentclass[a4paper,amsmath,amssymb,prl,floatfix,footinbib,twocolumn,preprintnumbers]{revtex4}

\usepackage[dvips]{graphicx}
\usepackage{dcolumn}
\usepackage{bm}
\usepackage{dsfont}
\usepackage{color}
\usepackage{ulem} 
\usepackage{url}
\usepackage[colorlinks=true ,citecolor=blue]{hyperref}
\usepackage{amsmath,amssymb}

\newcommand{\nn}{\nonumber}
\newcommand{\be}{\begin{equation}}
\newcommand{\ee}{\end{equation}}
\newcommand{\bea}{\begin{eqnarray}}
\newcommand{\eea}{\end{eqnarray}}
\newcommand{\mv}[1]{\langle #1\rangle}
\newcommand{\f}{\frac}
\newcommand{\ket}{\rangle}

\begin{document}

\title{Controlling coherence via tuning of the population imbalance in a bipartite optical lattice}

\author{M. Di Liberto$^1$, T. Comparin$^{1,2}$ , T. Kock$^3$, M. \"{O}lschl\"{a}ger$^3$, A. Hemmerich$^3$ and C. Morais Smith$^1$}
\affiliation{$^1$Institute for Theoretical Physics, Centre for Extreme Matter and Emergent Phenomena, Utrecht University, Leuvenlaan 4, 3584CE Utrecht, the Netherlands \\
$^2$Laboratoire de Physique Statistique, \'{E}cole Normale Sup\'{e}rieure, UPMC, Universit\'{e} Paris Diderot, CNRS, 24 rue Lhomond, 75005 Paris, France\\
$^3$ Institut f\"{u}r Laser-Physik, Universit\"{a}t Hamburg, LuruperChaussee 149 22761 Hamburg, Germany}

\date{\today}

\begin{abstract}
The control of transport properties is a key tool at the basis of many technologically relevant effects in condensed matter. The clean and precisely controlled environment of ultracold atoms in optical lattices allows one to prepare simplified but instructive models, which can help to better understand the underlying physical mechanisms. Here we show that by tuning a structural deformation of the unit cell in a bipartite optical lattice, one can induce a phase transition from a superfluid into various Mott insulating phases forming a shell structure in the superimposed harmonic trap. The Mott shells are identified via characteristic features in the visibility of Bragg maxima in momentum spectra. The experimental findings are explained by Gutzwiller mean-field and quantum Monte Carlo calculations. Our system bears similarities with the loss of coherence in cuprate superconductors, known to be associated with the doping induced buckling of the oxygen octahedra surrounding the copper sites.
\end{abstract}


\maketitle

\textit{Introduction}.
Rapid and precise control of transport properties are at the heart of many intriguing and technologically relevant effects in condensed matter. Small changes of some external parameter, for example, an electric or a magnetic field, may be used to significantly alter the mobility of electrons. Prominent examples are field effect transistors \cite{Bar:56} and  systems showing colossal magneto-resistance \cite{Jon:50}. Often, the control is achieved via structural changes of the unit cell, leading to an opening of a band gap. In iron-based superconductors, the variation of pressure is a well-known technique to control their transport properties \cite{Sun:12}. In certain high-$T_c$ superconductors, pulses of infrared radiation, which excite a mechanical vibration of the unit cell, can for short periods of time switch these systems into the superconducting state at temperatures where they actually are insulators \cite{Fau:11}. In La-based high-$T_c$ cuprates, the drastic reduction of $T_c$ at the doping value of $x = 1/8$, known as "the 1/8 mystery", is connected to a structural transition that changes the lattice unit cell \cite{Tra:13}. 

Ultra-cold atoms in optical lattices provide a particularly clean and well controlled experimental platform for exploring many-body lattice physics \cite{Lew:07}. Schemes for efficient manipulation of transport properties can be readily implemented and studied with great precision. In conventional optical lattices, tuning between a superfluid and a Mott insulating phase has been achieved by varying the overall lattice depth $V_0$, with the consequence of changing the height of the tunneling barriers and the on-site contact interaction energy \cite{Gre:02}. The equivalent is not easily possible in condensed matter systems, since the lattice depth is practically fixed. 

\begin{figure}[tbh]
\includegraphics[scale=0.35, angle=0, origin=c]{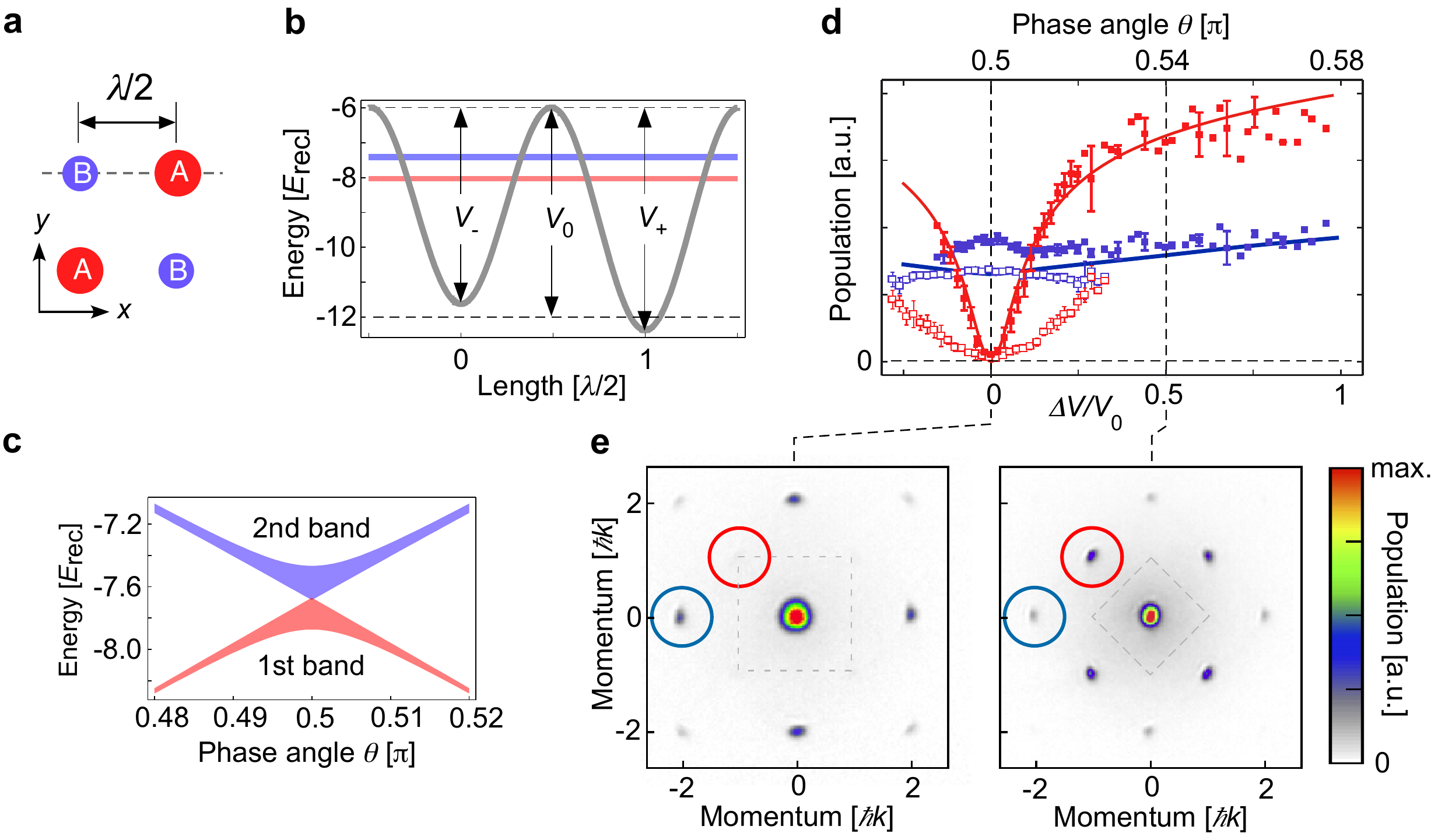}
\caption{Lattice potential. (a) Sketch of lattice geometry within the $xy$-plane. $\lambda = 1064$~nm denotes the wave length of the laser light.  In (b) the potential along the dashed trajectory in (a) is plotted for $\theta = 0.51\, \pi$ and $V_0 = 6 E_{\textrm{rec}}$ (thick grey line) with the first  and second bands represented, respectively, by the red (light gray) and blue (dark gray) horizontal bars. In (c) the first two bands are plotted versus $\theta$ for $V_0 = 6 E_{\textrm{rec}}$. On the bottom of (d), momentum spectra ($V_0 = 6 E_{\textrm{rec}}$, $V_{z,0}=0$) are shown with $\Delta V=0$ (left) and $\Delta V/V_0=0.5$ (right) with the respective FBZs imprinted as dashed rectangles. Above, the relative number of atoms (normalized to the total particle number) in the Bragg peaks indicated by red and blue squares are plotted versus $\Delta V / V_0$. The filled (open) squares are recorded for $V_{z,0}=0$ ($V_{z,0}=22 E_{\textrm{rec}}$). The solid lines are determined by a full band calculation (neglecting interaction) with no adjustable parameters. 
\label{fig_lattice}}
\end{figure}

\begin{figure}[hbt]
\includegraphics[width=0.85\columnwidth]{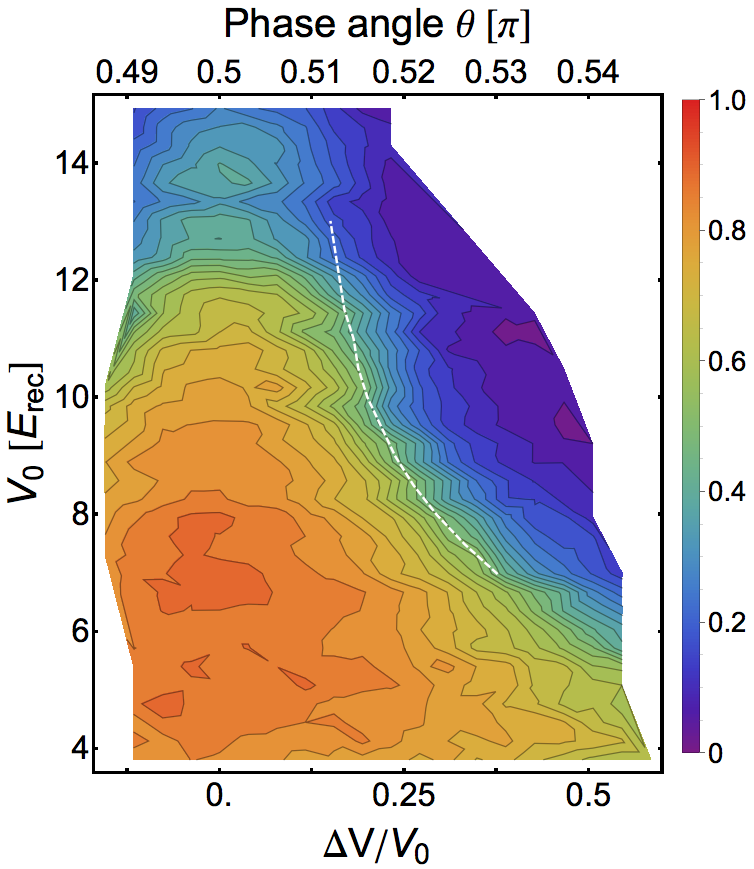}
\caption{Visibility measurements in the bipartite lattice. The visibility (parametrized by the color code shown on the right edge) is plotted as a function of the well depth parameter $V_0$ (measured in units of the recoil energy $E_{\rm rec}$) and the potential energy off-set difference $\Delta V$ between shallow and deep wells in the bipartite lattice. The dashed line corresponds to the theoretical calculation of the points where the fraction of particles $n_B=\sum_{i\in\mathcal{B}} n_i/N$ of the $\mathcal{B}$ sublattice vanishes ($n_B < 5.5\times 10^{-3}$).
\label{fig_visibility2D}}
\end{figure}

In this work, we present an ultracold atom paradigm, where tuning the system between a superfluid and a Mott insulator becomes possible via controlled distortion of the unit cell (see Fig.~\ref{fig_lattice}). This distortion acts to adjust the relative depth $\Delta V$ between two classes of sites (denoted by $\mathcal{A}$ and $\mathcal{B}$) forming the unit cell and allows us to drive a superfluid to Mott insulator transition without altering the average lattice depth. We can access a rich variety of Mott-insulating states with different integer populations of the $\mathcal{A}$- and $\mathcal{B}$-sites, which give rise to a shell structure in the finite harmonic trap potential, leading to characteristic features in the visibility of Bragg maxima in momentum spectra (see Fig.~\ref{fig_visibility2D}). We compare our observations with Quantum Monte Carlo (QMC) and Gutzwiller mean field calculations, thus obtaining a detailed quantitative understanding of the system. In the following, we first describe our experimental set-up; then, we theoretically investigate the behavior of the visibility along two different trajectories in Fig.~\ref{fig_visibility2D}: {\it i)} for fixed barrier height $V_0$, by varying $\Delta V$ (bipartite lattice), and {\it ii)} for $\Delta V = 0$ (monopartite lattice), by tuning the lattice depth $V_0$.  Although monopartite lattices have been previously studied in great detail, and QMC calculations have provided a good fitting of the visibility curve measured experimentally \cite{gerbier2008}, here we show more accurate data and argue that the main features of the curve can be understood in terms of a precise determination of the onset of new Mott lobes in the phase diagram. 

\textit{Description of the experimental set-up}. We prepare an optical lattice of $^{87}$Rb atoms 
using an interferometric lattice set-up \cite{Hem:91, Wir:11, Oel:11, Oel:13}. 
A two-dimensional (2D) optical potential is produced, comprising deep and shallow wells ($\mathcal{A}$ and $\mathcal{B}$ in Fig.~\ref{fig_lattice}(a)) arranged as the black and white fields of a chequerboard. In the $xy$-plane, the optical potential is given by $V(x,y) = -V_0 \left[ \cos^2 (k x) +  \cos^2 (k y) + 2 \cos(\theta)  \cos (k x)  \cos (k y)\right]$, with the tunable well depth parameter $V_0$ and the lattice distortion angle $\theta$. An additional lattice potential $V_{z}(z) \,\equiv - V_{z,0} \cos^2(k z)$ is applied along the $z$-direction. In order to study an effectively 2D scenario, $V_{z,0}$ is adjusted to $29 \,E_{\textrm{rec}}$, such that the motion in the $z$-direction is frozen out. Here, $k \equiv 2 \pi/ \lambda$, $E_{\textrm{rec}} \equiv \hbar^2 k^2/2m$, $m$ denotes the atomic mass, and $\lambda = 1064$~nm is the wave length of the lattice beams. Apart from the lattice, the atoms experience a nearly isotropic harmonic trap potential. Adjustment of $\theta$ permits controlled tuning of the effective well depths of the deep and shallow wells $V_{\pm} \equiv V_0\,(1\pm\cos(\theta))^2$ and their difference $\Delta V \equiv V_{+} - V_{-} = 4\, V_0 \cos(\theta)$ (see Fig.~\ref{fig_lattice}(b)). The effective mean well depth $\bar V_0 = (V_{+} + V_{-}) / 2 $ $= V_0 \left[1+\cos^2(\theta) \right]$ is only weakly dependent on $\theta$. For example, within the interval $0.46 < \theta / \pi < 0.54$ one has $ \cos^2(\theta) < 0.015$ and hence $\bar V_0 \approx V_0$. Tuning of $\theta$ significantly affects the effective band width, as shown in Fig.~\ref{fig_lattice} (c). At $\theta = \pi / 2$, the $\mathcal{A}$- and $\mathcal{B}$-wells become equal, which facilitates tunneling as compared to values $\theta \neq \pi / 2$, where the broad lowest band of the $\theta = \pi / 2$-lattice splits into two more narrow bands. 

We record momentum spectra, which comprise pronounced Bragg maxima with a visibility $\mathcal{V}$ (specified in the methods section) depending on the parameters $V_0$ and $\Delta V$. The distribution of Bragg peaks reflects the shape of the underlying first Brillouin zone (FBZ), which changes size and orientation as $\Delta V$ is detuned from zero. This is illustrated in Fig.~\ref{fig_lattice} (d), where two spectra recorded for $\Delta V=0$ (left) and $\Delta V /V_0=0.5$ (right) are shown. For $\Delta V=0$ (the special case of a monopartite square lattice), the increased size of the FBZ gives rise to destructive interference, such that the $\pm (1,\pm 1) \hbar k$-Bragg peaks indicated by the red circle vanish. As $\Delta V$ is detuned from zero, a corresponding imbalance of the $\mathcal{A}$- and $\mathcal{B}$-populations yields a retrieval of the $\pm (1,\pm 1) \hbar k$-Bragg peaks. This is shown for the case of approximately vanishing interaction energy per particle $U  \approx 0$ ($V_{z,0}=0$) by the filled red squares and for $U \approx  0.3 \,E_{\textrm{rec}}$ ($V_{z,0} = 22 \,E_{\textrm{rec}}$) by the open red squares, respectively. It is seen that the interaction energy significantly suppresses the formation of a population imbalance and corresponding $\pm (1,\pm 1) \hbar k$-Bragg peaks.

\textit{Model}. 
For low temperatures and for large lattice depths $V_0$, the system is described by the inhomogeneous Bose-Hubbard model \cite{fisher1989,jaksch1998} 
\begin{equation}
\label{eqBHconv}
H =
-J \sum_{\langle i,j \rangle} (a^\dag_i a_j + \textrm{h.c.})
-\sum_i \tilde{\mu}_i n_i
+\frac{U}{2} \sum_i n_i (n_i-1),
\end{equation}
where $J$ is the coefficient describing hopping between nearest-neighbor sites, $U$ accounts for the on-site repulsion, and $\tilde{\mu}_i$ is a local chemical potential, which  depends on the frequency $\omega$ of the trap and on the sublattice: $\tilde{\mu}_i=\mu_{A,B} - m\,\omega^2 {\bf r}_i^2 /2$. The ratio $U/J$ is a monotonously increasing function of $V_0/ E_{\textrm{rec}}$. \newline

{\it Bipartite Lattice $(\Delta V \ne 0)$}. The visibility measured for fixed $V_0$ as a function of $\Delta V$ (see Fig.~\ref{fig_visibility2D}) exhibits a region of rapid decrease. When the lattice barrier is large, e.g. $V_0= 12 \,E_{\rm rec}$, a modest detuning $\Delta V \sim 0.25\,V_0$ is able to completely destroy phase coherence with the consequence of a vanishing visibility. At smaller barrier heights, e.g. $V_0 = 6 E_{\rm rec}$, superfluidity remains robust up to significantly larger values of $\Delta V$. To explain this behavior, we performed a mean-field calculation using the Gutzwiller technique \cite{sheshadri1993} for the Bose-Hubbard model given by Eq. (\ref{eqBHconv}). The values of $J$ and $\Delta \mu = \mu_A - \mu_B$ have been estimated from the exact band structure and $U$ has been calculated within the harmonic approximation. The total number of particles has been fixed to $N=2\times 10^3$ and the trap frequency takes into account the waist of the laser beam (see Methods and Supplementary Information~\cite{supplementary}). We performed large-scale Gutzwiller calculations in presence of a trap, thus going beyond Local Density Approximation \cite{zakrzewski2005}.

\begin{figure}[hbt]
\includegraphics[width=0.85\columnwidth]{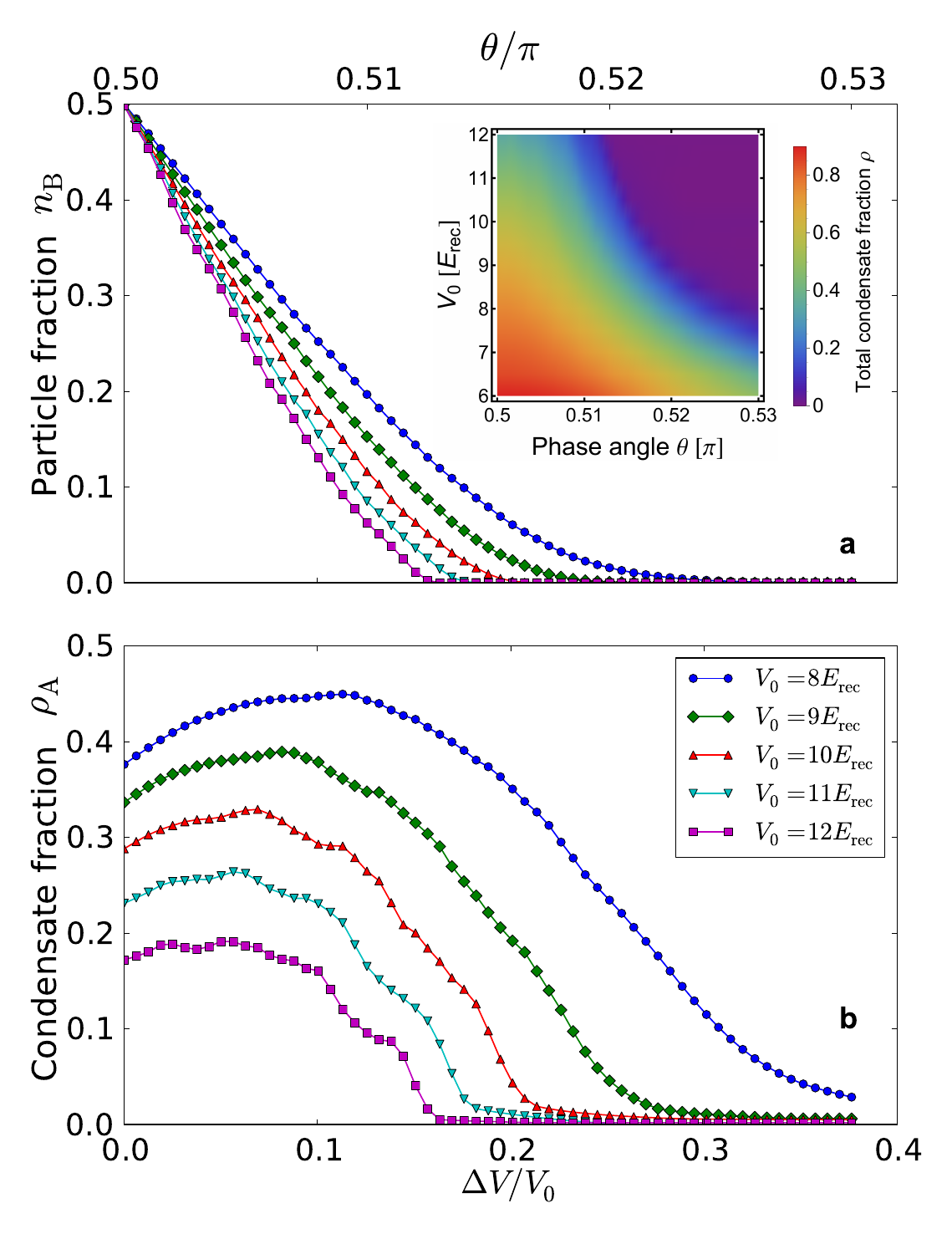}
\caption{Gutzwiller results in the trap. (a) Particle number fraction on the $\mathcal{B}$ sites ($n_B$). The inset shows the total condensed fraction $\rho = \sum_{i} \rho_i/N$. (b) Condensate fraction on the $\mathcal{A}$ sites ($\rho_A = \sum_{i\in \cal A} \rho_i/N$, where $\rho_i = |\phi_i |^2$, with $\phi_i$ the mean-field order parameter) as a function of $\Delta V$ for increasing values of $V_0$ and fixed total number of particles $N = 2\times 10^3$, calculated with the Gutzwiller ansatz. The inset shows the color code for both, the curves in (a) and (b).}
\label{numerical_simul}
\end{figure}

In Fig.~\ref{numerical_simul}(a), we show the evolution of the fraction of particles in the $\mathcal{B}$ sites (which we assumed to be the shallow wells). As $\Delta V$ increases, the number of bosons in the $\mathcal{B}$ sites decreases because of the excess potential energy required for their population. Within the tight-binding description, this is captured by the increased chemical potential difference between $\mathcal{A}$ and $\mathcal{B}$ sites as $\Delta V$ grows. Our calculations predict a critical value $\Delta V_c$ for which the population of the $\mathcal{B}$ sublattice vanishes. As shown in Fig.~\ref{numerical_simul}(a), $\Delta V_c$ becomes smaller as $V_0$ increases. This corresponds to the observation in the phase diagram shown in Fig.~9 of the Supplementary Information~\cite{supplementary} that the area covered by the Mott insulating regions with vanishing $\mathcal{B}$-populations (filling $g_B = 0$) increases as the hopping amplitude is reduced. The critical values $\Delta V_c$ for different values of $V_0$ are also shown in Fig.~\ref{fig_visibility2D} as a dashed white line on top of the experimental data for the visibility. This line consistently lies on experimental points corresponding to  constant visibility ($\mathcal{V}\approx 0.5$), where phase coherence is rapidly lost, and suggests the onset of a new regime.

 \begin{figure}[!htbp]	
 	\begin{center}
 	\includegraphics[width=0.85\columnwidth]{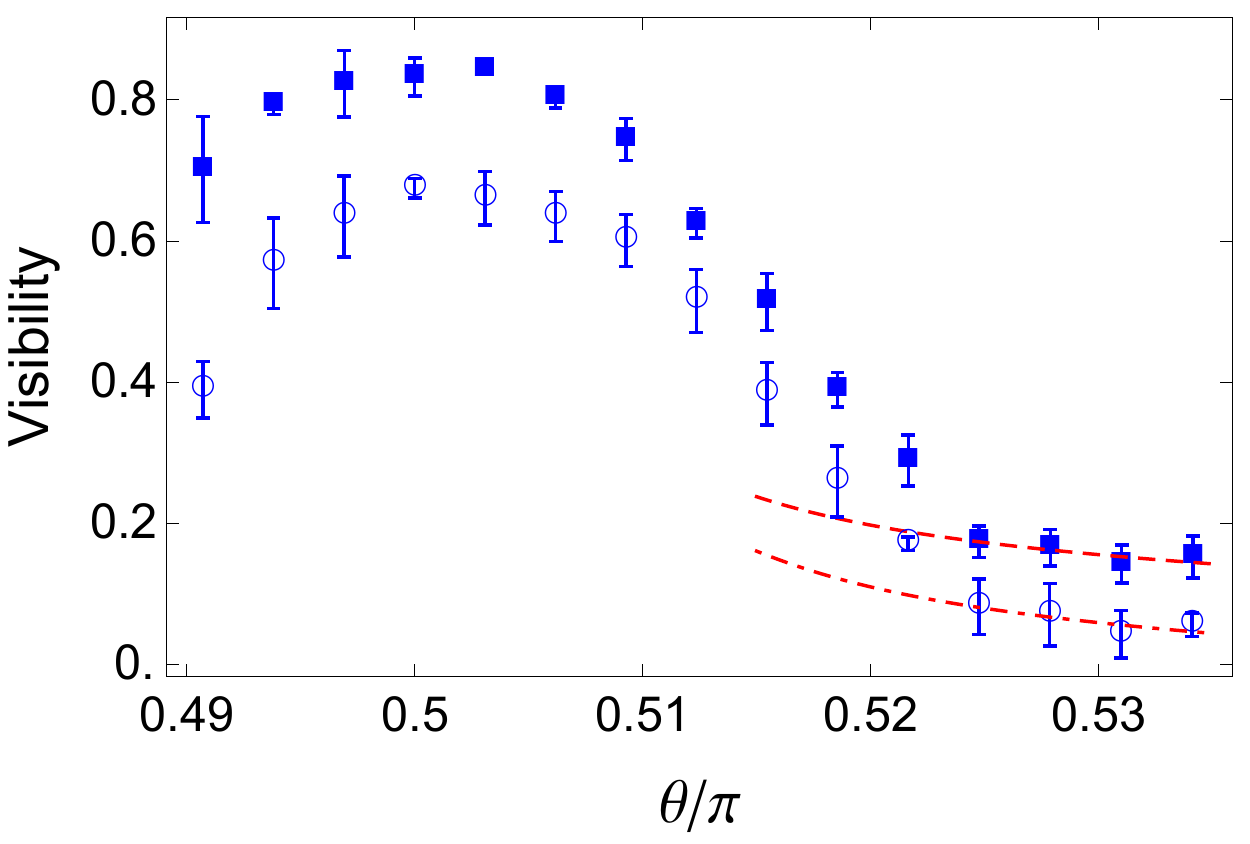}
 	\end{center}
 	\caption{Comparison of the measured visibility with the theory at large imbalance. The data shown are for $V_0 = 10.8\,E_{\rm{rec}}$ (squares) and $V_0 = 11.44\,E_{\rm{rec}}$ (circles). The red dashed (dash-dotted) line is obtained by fitting the last four data points with Eq.~(\ref{visib}) using the average filling $\bar g$ as a fitting parameter. We obtain respectively $\bar g=2.75\pm0.23$ and $\bar g=3.77\pm0.31$. The data for $V_0 = 10.8\,E_{\rm{rec}}$ are shifted along the vertical axis by $0.1$. The error bars represent the statistical variance of typically 4-5 independent measurements.}
 	\label{fig:datavisib0}
 \end{figure}

In Fig.~\ref{numerical_simul}(b) it is shown that, in addition to the population of the $\mathcal{B}$ sites, also the condensate fraction at the $\mathcal{A}$ sites approaches zero beyond the critical value $\Delta V_c$ (see the inset in Fig.~\ref{numerical_simul}(a) for the total condensed fraction); in this regime, the density profile displays only sharp concentric Mott shells of the form $(g_A, g_B) = (g,0)$ where the integer filling $g$ of the Mott regions can reach $g=4$ (see Supplementary Information).  This can be understood by considering that in the new regime where $\cal B$ sites are empty, the particles populating $\cal{A}$ sites can only delocalize (and thus establish phase coherence) by hopping through the intermediate $\cal{B}$ sites. Since these are second order processes, they are highly suppressed when $\Delta \mu$ is large enough and the system has to become an imbalanced Mott insulator.

In the new Mott-insulating regime, particle-hole pairs are responsible for a non-vanishing visibility, as in the conventional case in absence of imbalance \cite{gerbier2005prl}. By performing perturbation theory on top of the ideal Mott-insulating state $|MI\ket = \prod_{i\in A} | g\ket_i \prod_{j\in B} | 0\ket_j$, the ground state can therefore be written as (see Supplementary Note 4)
\bea
|\psi_{\rm G}\ket &=& \left(1 - \f{J^2}{2\Delta^2} \right)|MI \ket -\f{J}{\Delta}\sum_{\mv{i,j}} a^\dag_i a_j |MI \ket \\ 
&& -\f{2J^2}{U\Delta}\sum_{\mv{i,j}_{{\rm A}}} a^\dag_i a_j |MI \ket  -\f{J^2}{U\Delta}\sum_{\mv{\mv{i,j}}_{{\rm A}}} a^\dag_i a_j |MI \ket\,, \nn
\eea
where $\Delta \equiv U (g-1) + \Delta \mu$.  The first term is simply the unperturbed term with a wave function renormalization, whereas the linear term in $J$ describes particle-hole pairs with the particle sitting on the ${A}$ site and the hole in the neighbor ${B}$ site, or vice-versa. The last two terms are second order processes that involve intermediate $B$ sites and describe particle-hole pairs within the $A$ sublattice only. This ground state leads to the visibility 
\be
\label{visib}
\mathcal{V} = c_1 J / \Delta + c_2 J^2 /U \Delta  + c_3 J^2 / \Delta^2, 
\ee
where $c_1 = - 2 (\bar{g}+1) (1-r_1)$, $c_2 =  4(\bar{g}+1)(2r_1 + r_2 - 3)$, $c_3 = - 4(\bar{g}+1)^2(r_1+3)(1-r_1)$, with $r_1\equiv \cos(\sqrt 2 \pi) \approx -0.266$ and $r_2\equiv \cos(\sqrt 8 \pi) \approx -0.858$. By using the average filling $\bar{g}$ in the trap as a fitting parameter, we found that the theoretical visibility curve compares reasonably well with the experimental data both in magnitude and scaling behavior, with an average filling of the order $\bar{g}\approx 3$ (see Fig.~\ref{fig:datavisib0}). A perturbative description of the visibility data for large $\theta$ by means of Eq.~(\ref{visib}) is only possible in a window $V_0 \approx 11 \pm 1 \, E_{\rm{rec}}$, where sufficient data points are available in the low visibility tail with values of the visibility large enough to be measured with sufficient precision to allow fitting. (see Supplementary Information \cite{supplementary}). \\

\begin{figure}[tbh]
\includegraphics[width=1\columnwidth]{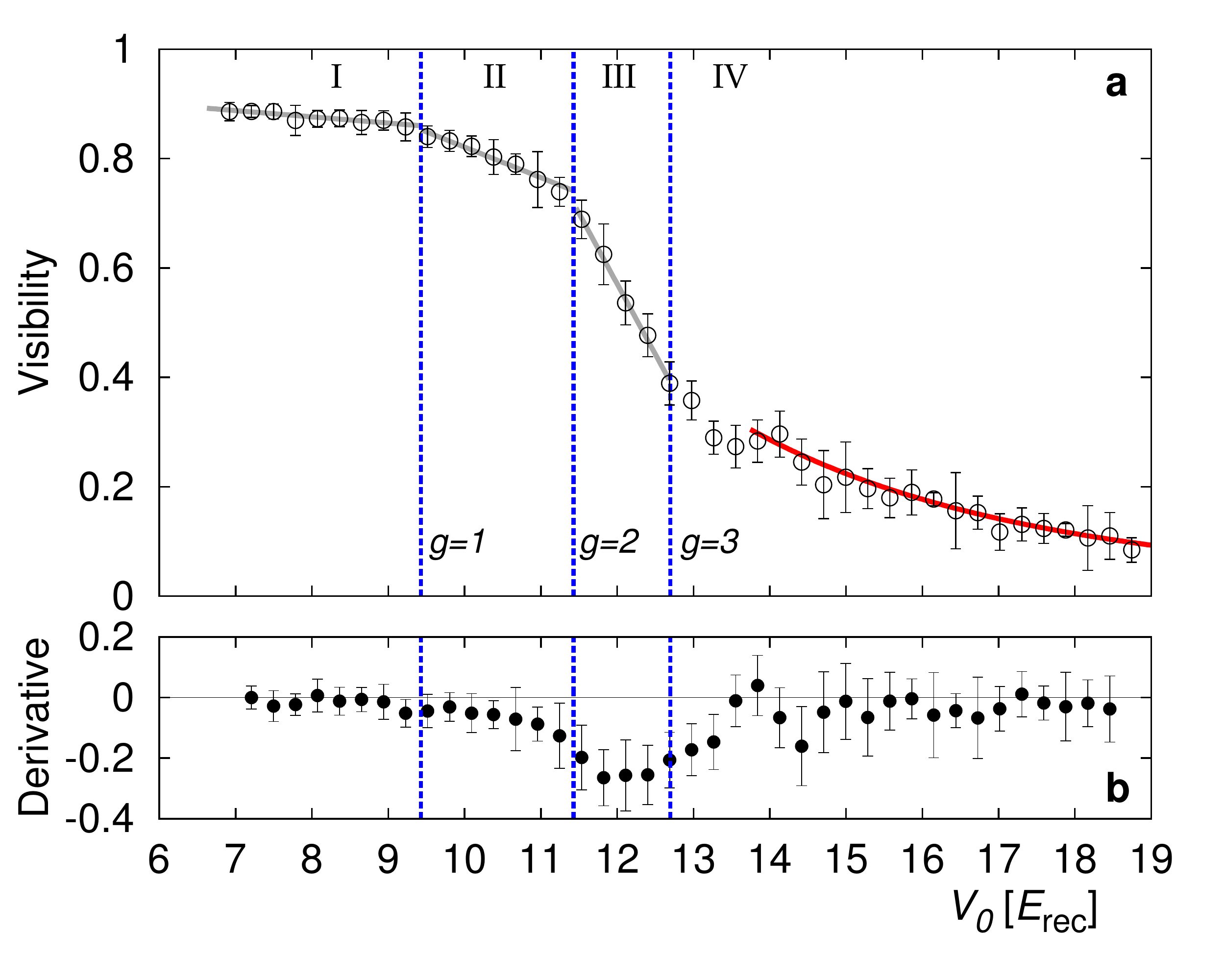}
\caption{Visibility measurement in the monopartite lattice. (a) Visibility of $^{87}$Rb, plotted as a function of the well depth $V_0$, for $\Delta V = 0$ and $V_{z,0} = 29\,E_{\textrm{rec}}$. Vertical dashed lines: values of $V_0/E_{\textrm{rec}}$ corresponding to the tips of the Mott lobes with different filling $g$, as computed through QMC (see \cite{supplementary}). Grey solid lines in regions I - III are a guide to the eyes, whereas the red line in region IV displays a fit to the function $A (U/zJ)^{\alpha}$ with $A=4.0 \pm 0.7$ and $\alpha=-1.00\pm0.06$, showing good agreement with the theoretical prediction \cite{gerbier2005prl}. (b) Numerical derivative of the visibility data; vertical lines as in (a).
\label{fig_visib_conventional}}
\end{figure}

\textit{Monopartite lattice $(\Delta V = 0)$}.
Adjustment of $\Delta V = 0$ produces the special case of a conventional monopartite square lattice, extensively studied in the literature during the past decade \cite{Gre:02, gerbier2005prl, gerbier2005pra, jimenez-garcia2010}. Experiments in 3D cubic lattices have suggested that the formation of Mott shells within the external trap could be associated to the appearance of kinks in the visibility \cite{gerbier2005prl,gerbier2005pra}, whereas experiments in 2D triangular lattices have rather detected an instantaneous decrease \cite{becker2010}. Arguable attempts were made to interprete small irregularities in the observed visibility in this respect. On the theoretical front, a QMC study of the 1D trapped Bose-Hubbard model \cite{sengupta2005} has shown the appearance of kinks in ${\cal V}$ as a function of $U/J$. Unfortunately, this study, employing a trap curvature proportional to $J$ rather than $V_0$, appears to have limited relevance for experiments. More realistic QMC simulations of 2D and 3D confined systems have been able to quantitatively describe the momentum distribution \cite{kashurnikov2002} and the experimental visibility \cite{pollet2008, gerbier2008}, however with no indications for distinct features  associated to Mott shells.

To clarify this long standing discussion, we have recorded the visibility of Fig.~\ref{fig_visibility2D} along the $\Delta V = 0$ trajectory versus $V_0$ with increased resolution in Fig.~\ref{fig_visib_conventional}. 
Guided by an inhomogeneous mean-field calculation indicating that the local filling $g$ is lower than 4, we computed the critical $J/U$ values for the tips of Mott lobes with $g = 1, 2$ and 3, by making use of the worm algorithm as implemented in the ALPS libraries \cite{capogrosso-sansone2008,albuquerque2007,bauer2011}. Superimposed upon the experimental data, we mark in Fig.~\ref{fig_visib_conventional} with (blue) dashed lines the values of $V_0/E_{\textrm{rec}}$ corresponding to the values of $J/U$ at the tip of the Mott lobes obtained by QMC. As $V_0$ is increased in Fig.~\ref{fig_visib_conventional}, four different regimes are crossed. For small values of $V_0$ (regime I), most of the system is in a superfluid phase. Increasing $V_0$ yields only little loss of coherence due to increasing depletion, and hence the visibility remains nearly constant. When the first Mott ring with $g=1$ particle per site is formed, the system enters regime II, where the visibility decreases slowly but notably as the $g=1$-Mott shell grows. When the second Mott-insulating ring with $g=2$ arises (regime III), a sharp drop of the visibility occurs indicating a significantly increased growth of the Mott-insulating part of the system with $V_0$. Finally, when the third Mott ring with $g=3$ forms or closes in the center of the trap, only a small superfluid fraction remains in the system, such that the visibility cannot further rapidly decrease with $V_0$ (regime IV), i.e., a quasi-plateau arises in Fig.~\ref{fig_visib_conventional}. The red solid line shows that for large $V_0$ the visibility acquires a $(U/J)^{-1}$ dependence, in agreement with a result obtained by first-order perturbation theory in $J/U$ \cite{gerbier2005prl}.

Several conclusions can be drawn from our experimental and theoretical investigations: for monopartite lattices the visibility comprises characteristic signatures, which can be connected to the position of the tips of the Mott-insulator lobes in a $\mu/U$ versus $J/U$ phase diagram calculated by QMC. Mean-field calculations are insufficient, even when the inhomogeneity due to the trap is taken into account. Deforming the unit cell of a bipartite lattice is a means to efficiently tune a transition from a superfluid to a Mott-insulating state. The visibility displays distinct regions with explicitly different slope, as a function of the detuning between the ${\cal A}$ and ${\cal B}$ sublattices. A pronounced loss of coherence occurs at the critical value of the detuning $\Delta V_c$, at which the population of the shallow wells vanish. Our work may shed some light also on the behavior of condensed-matter systems, where loss of phase coherence occurs due to a structural modification of the lattice. For example, in La$_{2-x}$Ba$_x$CuO$_4$ high-$T_c$ cuprate, superconductivity is weakened at the structural transition from a low-temperature orthorhombic (LTO) into a low-temperature tetragonal (LTT) phase \cite{1/8}. The same occurs for La$_{2-x-y}$Nd$_y$Sr$_x$CuO$_4$ \cite{Tra:13}. This structural transition corresponds to a buckling of the oxygen octahedra surrounding the copper sites, which changes the nature of the copper-oxygen lattice unit cell \cite{1/8}. The critical buckling angle $\theta_c = 3.6 \deg$ for the destruction of superconductivity \cite{Kampf} bears similarities with the critical deformation angle $\theta_c$ (or equivalently $\Delta V_c$) found here. Most of the present theoretical studies of high-$T_c$ superconductivity concentrate only on the copper lattice. We hope that our results will inspire further investigations of the specific role played by the oxygen lattice, and its importance in preserving phase coherence.
\\ \\
\textbf{ Acknowledgments.}
This work was partially supported by the Netherlands Organization for Scientific Research (NWO), by the German Research Foundation DFG-(He2334/14-1, SFB 925), and the Hamburg centre of ultrafast imaging (CUI). A. H. and C.M.S acknowledge support by NSF-PHYS-1066293 and the hospitality of the Aspen Center for Physics. We are grateful to Peter Barmettler, Matthias Troyer and Juan Carrasquilla for helpful discussions. 
\\ \\
\textbf{ Methods.}
Our experimental procedure begins with the production of a nearly pure Bose-Einstein condensate of typically $5 \times 10^{4}$ rubidium atoms ($^{87}$Rb) in the $F=2, m_F = 2$ state confined in a nearly isotropic magnetic trap with about 30 Hz trap frequency. The adjusted values of the lattice depth $V_0$ are determined with a precision of about 2 percent by carefully measuring the resonance frequencies with respect to excitations into the third band along the $x$- and $y$-directions. The adjustment of $\theta$ is achieved with a precision exceeding $\pi/300$ by an active stabilization with about $10\,$kHz bandwidth. In a typical experimental run, the lattice potentials $V(x,y)$ and $V_{z}(z)$ are increased to the desired values by an exponential ramp of 160 ms duration. After holding the atoms in the lattice for 20 ms, momentum spectra are obtained by rapidly ($< 1\,\mu$s) extinguishing the lattice and trap potentials, permitting a free expansion of the atomic sample during 30 ms, and subsequently recording an absorption image. The magnetic trap  and the finite Gaussian profile of the lattice beams (beam radius = 100 $\mu$m) give rise to a combined trap potential. For $V_{z,0} = 29 \,E_{\textrm{rec}}$ and $V_{0} = 18 \,E_{\textrm{rec}}$ this yields trap frequencies of 73 Hz in the $xy$-plane and 65 Hz along the $z$-direction. The observed momentum spectra comprise pronounced Bragg maxima with a visibility depending on the parameters $V_{0}$ and $\Delta V$. These spectra are analyzed by counting the atoms ($n_{d,0}$) in a disk with 5 pixel radius around some higher order Bragg peak and within a disk of the same radius but rotated with respect to the origin by $45^{\circ}$ ($n_{d,45}$). The visibility is obtained as ${\mathcal V} = (n_{d,0} - n_{d,45})/(n_{d,0} + n_{d,45})$ \cite{gerbier2005prl}.


\clearpage
\onecolumngrid

\begin{center}
{\Large SUPPLEMENTARY INFORMATION}
\end{center}

\subsection{I. Band-structure and tight-binding model}

We employ an optical lattice with two classes of wells (denoted as $\mathcal{A}$ and $\mathcal{B}$) arranged as the black and white fields of a chequerboard. The optical potential is
\be
\label{potential}
V(x,y) = -V_0 \left[ \cos^2 (k x) +  \cos^2 (k y) + 2 \cos\theta  \cos (k x)  \cos (k y)\right]\,
\ee
with the tunable well depth parameter $V_0$ and the lattice distortion angle $\theta$. Adjustment of $\theta$ permits controlled tuning of the well depth difference $\Delta V \equiv 4\, V_0 \cos(\theta)$ between $\mathcal{A}$ and $\mathcal{B}$ sites. In the special case $\theta=\pi/2$ ($\Delta V=0$) both types of sites are equivalent and hence a monopartite lattice arises, while in general the lattice is bipartite. Using the potential of Eq.~(\ref{potential}), we have numerically solved the Schr\"{o}dinger equation for the single particle problem to obtain the exact band structure, including 14 bands in the plane-wave matrix representation of the Hamiltonian \cite{paul2013}.

\begin{figure}[htb]
\centering
\includegraphics[width=0.3\textwidth]{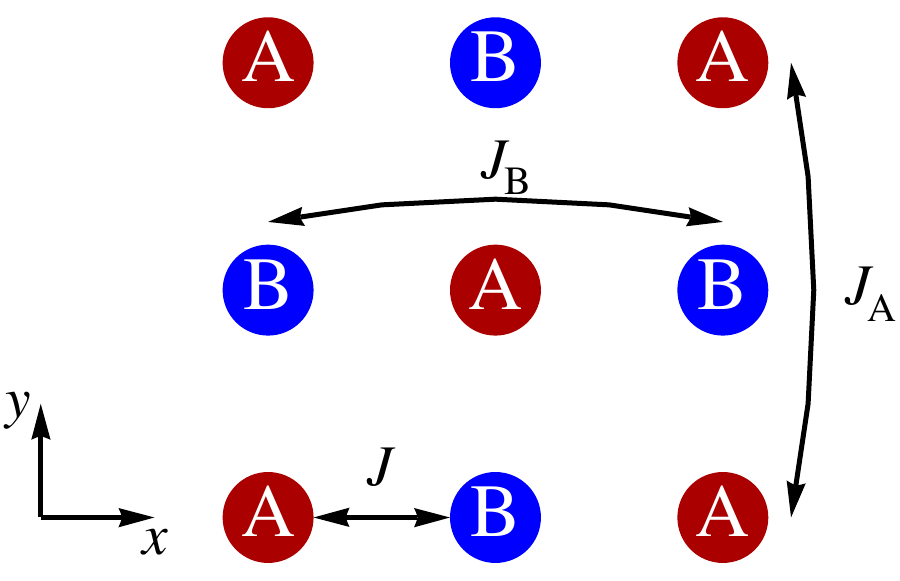}
\caption{Hopping processes included in the tight-binding model.
\label{fig:lattice_tight_binding}}
\end{figure}

\begin{figure}[htb]
\centering
\includegraphics[width=0.3\textwidth]{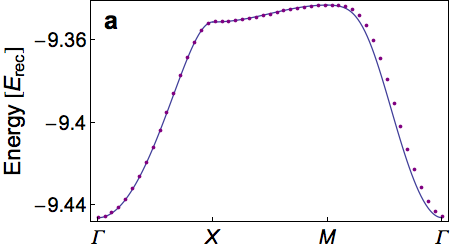}
\includegraphics[width=0.285\textwidth]{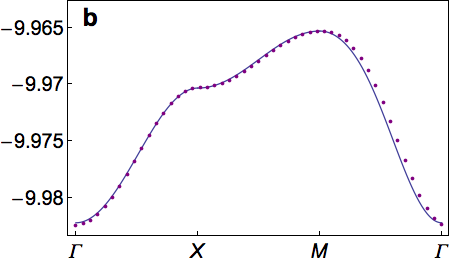}
\includegraphics[width=0.29\textwidth]{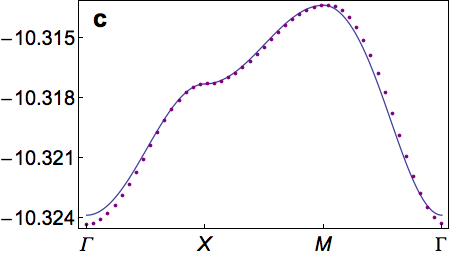}
\caption{Lowest band in the bipartite optical potential. The lowest energy band for $V_0 = 7~E_{\rm rec}$ and (a) $\theta=0.502\,\pi$, (b) $\theta=0.52\,\pi$, and (c) $\theta=0.53~\pi$ is plotted versus $k$, along the high-symmetry lines of the first Brillouin zone. The dots are the results of the exact diagonalization, the solid line is calculated according to the tight-binding Hamiltonian (\ref{model}) with parameter values determined according to Eq.~(\ref{parameters}).
\label{fig:bands}}
\end{figure}

The single-particle problem is reformulated in terms of a tight-binding model Hamiltonian,
\be
\label{model}
H = -J \sum_{\mv{i,j}} (a^\dag_i a_j + \textrm{h.c.}) -J_A  \sum_{\mv{i,j}_\mathcal{A}} (a^\dag_i a_j + \textrm{h.c.}) -J_B  \sum_{\mv{i,j}_\mathcal{B}} (a^\dag_i a_j + \textrm{h.c.}) +E_A \sum_{i\in \mathcal{A}} n_i +E_B \sum_{i\in \mathcal{B}} n_i\,,
\ee
where $J$ is the hopping between neighboring sites of different sublattices, $J_A$ ($J_B$) is the hopping coefficient between neighboring sites of sublattice $\mathcal{A}$ ($\mathcal{B}$) (see Fig.~\ref{fig:lattice_tight_binding}), and $E_A (E_B)$ is the on-site energy of sites belonging to sublattice $\mathcal{A}$ ($\mathcal{B}$). We neglect the $\mathcal{A}\rightarrow \mathcal{A}$ hopping (henceforth indicated as $J_A'$) along the diagonal lines of the lattice (same for $\mathcal{B}\rightarrow \mathcal{B}$), because for the monopartite lattice  $(\theta =\pi/2$ or $\Delta V = 0)$ these hopping coefficients are exactly zero as a consequence of the symmetry of the Wannier functions. For sufficiently small deviations from $\theta = \pi/2$, we expect that these coefficients are still negligible compared to $J_A$ or $J_B$; this assumption is supported by the full band structure calculation. For $\theta \gtrsim 0.53 \,\pi$, this assumption becomes less reliable (see Fig.~\ref{fig:bands}).

\begin{figure}[htb]
\centering
\includegraphics[width=0.4\textwidth]{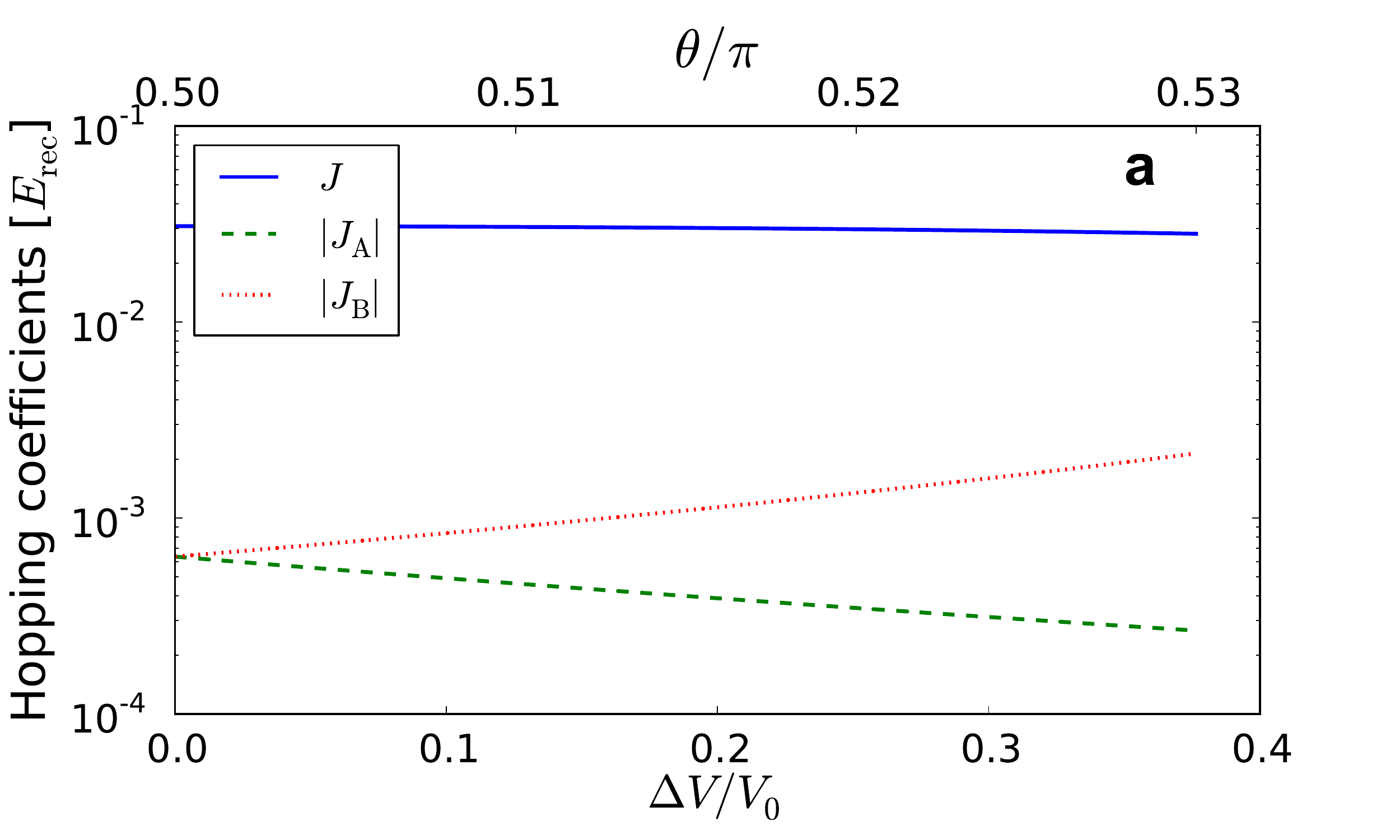}
\includegraphics[width=0.4\textwidth]{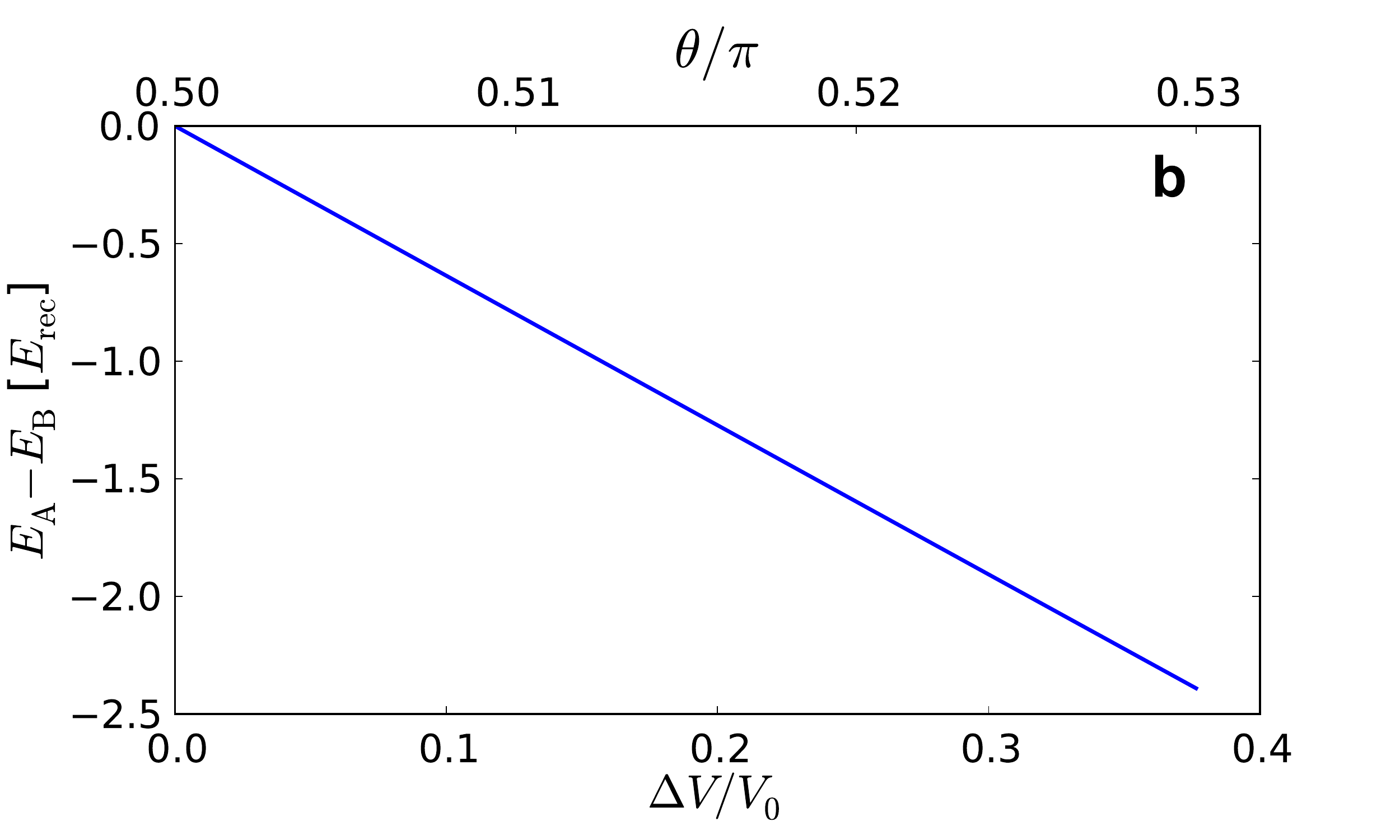}
\caption{Tight-binding parameters. (a) Hopping coefficients and (b) energy difference $E_A - E_B$ versus $\Delta V$ (or equivalently $\theta$) according to Eq.~(\ref{parameters}) for $V_0 = 8\,E_{\rm rec}$.
\label{hoppcoeff}}
\end{figure}

By diagonalizing the Hamiltonian in Eq.~(\ref{model}) in momentum space and taking the lattice constant to unity, an analytic expression for the corresponding band structure (depending on the parameters $E_A,E_B,J,J_A,J_B$) can be derived, 
\be
H(\bf{k}) = \begin{pmatrix}
E_A - 4 J_A \cos(2 k_x) \cos(2 k_y ) &  -4 J \cos(k_x ) \cos(k_y) \\
-4 J \cos(k_x) \cos(k_y ) & E_B - 4 J_B \cos(2 k_x) \cos(2 k_y)
\end{pmatrix}\,.
\ee 
When $\theta$ is tuned away from zero a gap opens, splitting the lowest band. We denote the two resulting bands by ``1'' and ``2'', with the corresponding energies $E_1(k_x,k_y)$ and $E_2(k_x,k_y)$. It is straightforward to verify that
\bea
\label{parameters}
E_A &= &E_1(\pi/2,\pi/4)\,,\nn\\
E_B &= &E_2(\pi/2,\pi/4)\,,\nn\\
J &=& \f 1 4 \sqrt{\left( E_1(\pi/4,\pi/4) - E_2(\pi/4,\pi/4) \right)^2 - (E_A - E_B)^2}\,,\nn\\
J_A &=& \f 1 8 ( E_1(\pi/2,0) - E_1(\pi/2,\pi/2) )\,,\nn\\
J_B &=& \f 1 8 ( E_2(\pi/2,0) - E_2(\pi/2,\pi/2) )\,.
\eea
In order to determine the parameters of the model Hamiltonian (\ref{model}), instead of calculating Wannier functions, we use these equations to adjust the tight-binding bands to the exact band structure calculation, finding reasonable agreement up to $\theta = 0.53\,\pi$, as shown in Fig.~\ref{fig:bands}. The resulting values of the hopping coefficients and the energy difference $E_A - E_B$ are plotted in Fig.~\ref{hoppcoeff}. Since $|J_A|$ and $|J_B|$ are nearly two orders of magnitude smaller than $J$, we will neglect them in what follows, as long as $J \ne 0$.  

\subsection{II. Mean-field phase diagram of the bipartite lattice model}

\begin{figure}[htb]
\centering
\includegraphics[width=0.5\textwidth]{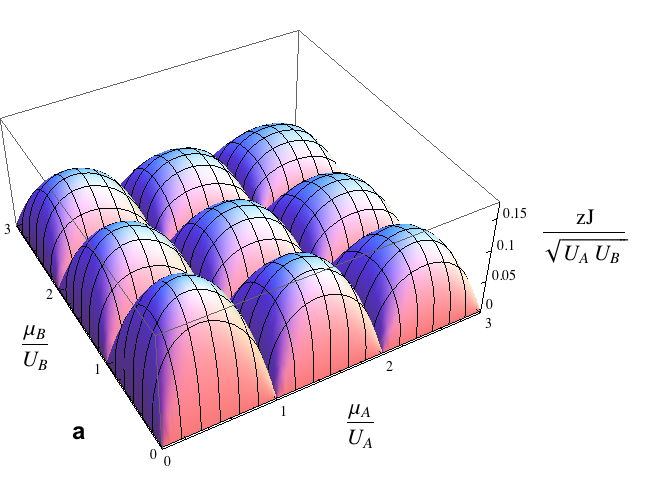}
\includegraphics[width=0.3\textwidth]{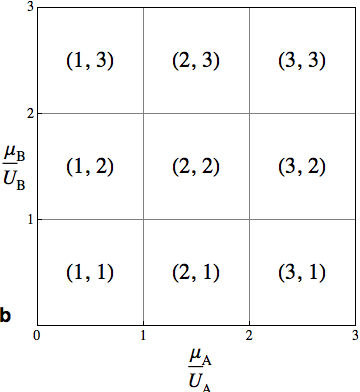}
\caption{Mean-field phase diagram in the bipartite lattice. (a) Phase boundaries: inside each lobe there is a Mott insulator phase with occupations that can differ in the two sublattices according to the chemical potentials, above the lobes the gas is superfluid. (b) Configurations of the occupation numbers $(g_A, g_B) $ inside each lobe.
\label{fig:phdiag}}
\end{figure}

In this section, we summarize the calculation of the mean-field phase diagram of the bosonic model
\bea
\label{modelbp}
H = -J \sum_{\mv{i,j}} (a^\dag_i a_j + \textrm{h.c.}) - \mu_A \sum_{i\in \mathcal{A}} n_i -  \mu_B \sum_{i\in \mathcal{B}} n_i+\f{U_A}{2} \sum_{i\in \mathcal{A}} n_i (n_i-1) + \f{U_B}{2} \sum_{i\in \mathcal{B}} n_i (n_i-1). 
\eea
We restrict ourselves to nearest-neighbor hopping coefficients, which are the only relevant ones, as shown in Fig.~\ref{hoppcoeff}. The Hamiltonian in Eq.~(\ref{modelbp}) describes a bipartite lattice, in which one allows for different densities in the two sublattices. A similar problem has been discussed also in Ref.~\cite{chen2010}. Extending a standard approach \cite{sheshadri1993,vanoosten2001}, we apply a mean-field decoupling of the hopping term [first term in Eq.~(\ref{modelbp})],
\be
a_{i(A)} \rightarrow \psi_A + \delta a_{i(A)}\,, \qquad a_{i(B)} \rightarrow \psi_B+ \delta a_{i(B)} 
\ee
with the order parameters $\psi_{A,B}\equiv \mv{a_{i(A,B)}}$ and the fluctuations  $\delta a_{i(A,B)}$. Neglecting the second order fluctuations of the fields, one finds
\be
H_J \simeq -4J  \sum_{i\in \mathcal{A}}\left(\psi_B a_i + \psi_B^* a^\dag_i \right) -4J  \sum_{i\in \mathcal{B}}\left(\psi_A a_i + \psi_A^* a^\dag_i \right) + 4N_A J (\psi^*_A \psi_B + \psi_A \psi^*_B)\equiv H_{J0} + 4N_A J (\psi^*_A \psi_B + \psi_A \psi^*_B)\,,
\label{HJ0}
\ee
where $N_A$ denotes the number of sites in the sublattice $\mathcal{A}$. We use $H_{J0}$ as a perturbation to the interaction part of the Hamiltonian (\ref{modelbp}), and neglect for the moment the irrelevant constant shift given by the last term in Eq.~(\ref{HJ0}). Since $H_{J0}$ is local, the total Hamiltonian contains only local terms and we can apply perturbation theory in each unit cell. The unperturbed Hamiltonian $H(J=0)$ is diagonal with respect to the number operators and, hence, the eigenstates of $H(J=0)$ in each unit cell are $| n_A, n_B \rangle$, where $n_A$ and $n_B$ are the occupation numbers of the sites $\mathcal{A}$ and $\mathcal{B}$, respectively. The energy per unit cell is given by 
\be
E(n_A, n_B) = \f{U_A}{2} n_A (n_A - 1) + \f{U_B}{2} n_B (n_B - 1) - \mu_A n_A - \mu_B n_B\,.
\ee
The ground state corresponds to occupations $g_A$ and $g_B$ determined by the relations $U_\nu (g_\nu-1) < \mu_\nu < U_\nu g_\nu$, with $\nu = A,\,B$. The first order contribution of the perturbation $H_{J0}$ vanishes because $H_{J0}$ does not conserve the number of particles, whereas the  second order is found to be 
\bea
E^{(2)} &=& \sum_{(n_A,n_B)\neq(g_A,g_B)} \f{| \langle g_A, g_B | H_{J0} | n_A,n_B\rangle |^2}{E(g_A,g_B) - E(n_A,n_B)} \nn\\
&=& (4J)^2 \left[ \f{|\psi_B|^2 g_A}{U_A(g_A - 1) - \mu_A} +   \f{|\psi_B|^2 (g_A+1)}{\mu_A - U_A g_A} + \f{|\psi_A|^2 g_B}{U_B(g_B - 1) - \mu_B} +   \f{|\psi_A|^2 (g_B+1)}{\mu_B - U_B g_B}  \right]\,.
\eea
Including the previously ignored constant shift and using the fact that at zero temperature the calculated energy is the same as the Helmholtz free energy $F$, we can write
\be
F[\psi_A, \psi_B] = F^{(0)} + \sum_{\mu,\nu=A,B}\psi^*_\mu M_{\mu\nu}\psi_\nu
\ee
with 
\be
F^{(0)} = \f{U_A}{2}g_A(g_A-1) + \f{U_B}{2}g_B(g_B-1) - \mu_A g_A - \mu_B g_B
\ee
and
\be
\textbf{M} = \begin{pmatrix}
\left(\f{g_B}{U_B(g_B - 1) - \mu_B} + \f{g_B+1}{\mu_B - U_B g_B} \right)J^2 z^2 & zJ \\
zJ & \left(\f{g_A}{U_A(g_A - 1) - \mu_A} + \f{g_A+1}{\mu_A - U_A g_A} \right)J^2 z^2
\end{pmatrix}\,.
\ee 
Here, $z=2d$ is the coordination number of the lattice; in our case $d=2$ and $z=4$. According to the (generalized) Landau criterion for continuous phase transitions, the phase boundaries are given by the condition $\rm{Det}[\textbf{M}] = 0$. In the phase diagram shown in Fig.~\ref{fig:phdiag}(a), one observes a series of lobes corresponding to Mott-insulator phases with occupation numbers that can vary in the two sublattices according to the value of the chemical potentials (see also Fig.~\ref{fig:phdiag}(b), where the $(g_A,g_B)$  filling of the Mott lobes is explicitly given). Outside the lobes the system is superfluid.

\subsection{III. Effect of the trap}

\begin{figure}[htb]
\centering
\includegraphics[width=0.8\textwidth]{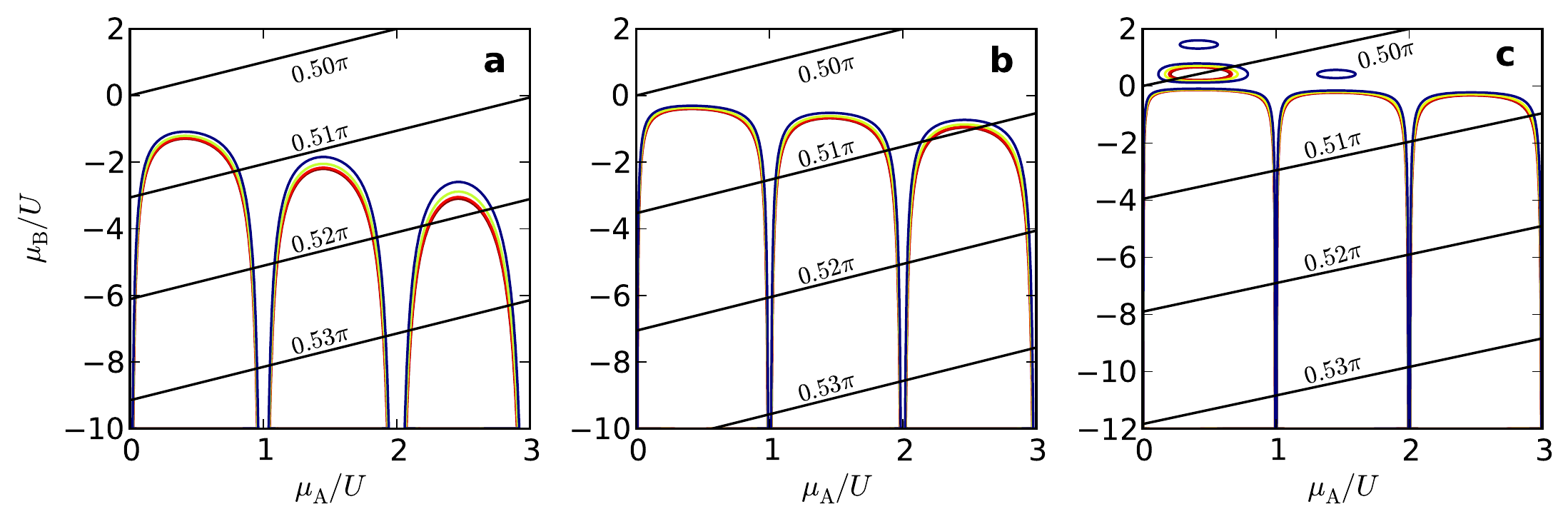}
\caption{Sections through the mean-field phase diagram of the bipartite lattice. For (a) $V_0 = 8 E_{\rm rec}$, (b) $V_0 = 10 E_{\rm rec}$, and (c) $V_0 = 12 E_{\rm rec}$, sections through the phase diagram in Fig.~\ref{fig:phdiag} at fixed values of $J/U$ are shown. In each panel, the small change of the Mott lobe boundaries with $\theta$ is indicated by contours of different colors; the largest lobe corresponds to the largest value of $\theta$, i.e. the lowest value of $J/U$ (see Sec.~I). The diagonal lines are given by Eq.~(\ref{chemcons}), for different values of $\theta$ (or equivalently, different values of $\Delta V/V_0$).
\label{slices}}
\end{figure}

We now discuss the effect of the additional harmonic trap potential with respect to the interpretation of the Gutzwiller calculations performed for a homogeneous system. We set $U_A = U_B=U$, which is a very good approximation for $\theta \lesssim 0.53~\pi$. In Fig.~\ref{slices}, horizontal sections through the mean-field phase diagram are plotted for fixed values of $V_0$. The lobes for $\mu_B<0$ correspond to Mott phases with occupations $(g_A,g_B)=(g,0)$, with $g$ integer. For different values of $\theta$, we also plot the lines $\mathcal{L}(\theta)$ given by
\be
\label{chemcons}
\mu_B -\mu_A = \Delta\mu(\theta) \, ,
\ee
where $\Delta\mu(\theta) = E_A - E_B$ is the difference of the local energies $E_A$ and $E_B$ determined through Eq.~(\ref{parameters}). According to the local density approximation, one can define a local chemical potential with a maximal value in the center of the trap fixed by the total particle number, which decreases towards the edge of the trap. Hence, the phases encountered locally along a radial path pointing outwards from the trap center are given by the homogeneous phase diagram, when following the lines $\mathcal{L}(\theta)$ towards decreasing values of $\mu_A/U$. The lines $\mathcal{L}(\theta)$ shift to large, negative values of $\mu_B/U$ as $\theta$ increases. As discussed in the main text, this means that the population of the $\mathcal{B}$ sites decreases and eventually vanishes. Hence, the density profile evolves into a wedding cake structure where only the $\mathcal{A}$ sites are populated, i.e., most atoms contribute to pure $\mathcal{A}$-site Mott shells $(g,0)$ separated by narrow superfluid films, also with negligible $\mathcal{B}$ population. The plot also shows that for increasing $V_0$ the Mott lobes $(g,0)$ cover an increasing area in the phase diagram, while at the same time the lines $\mathcal{L}(\theta)$ shift towards lower values of $\mu_B/U$. This explains why the value of $\Delta V_c$, at which one observes a sudden loss of the visibility, reduces when $V_0$ is increased.

\begin{figure}[!htbp]	
 	\begin{center}
 	\includegraphics[width=0.45\textwidth]{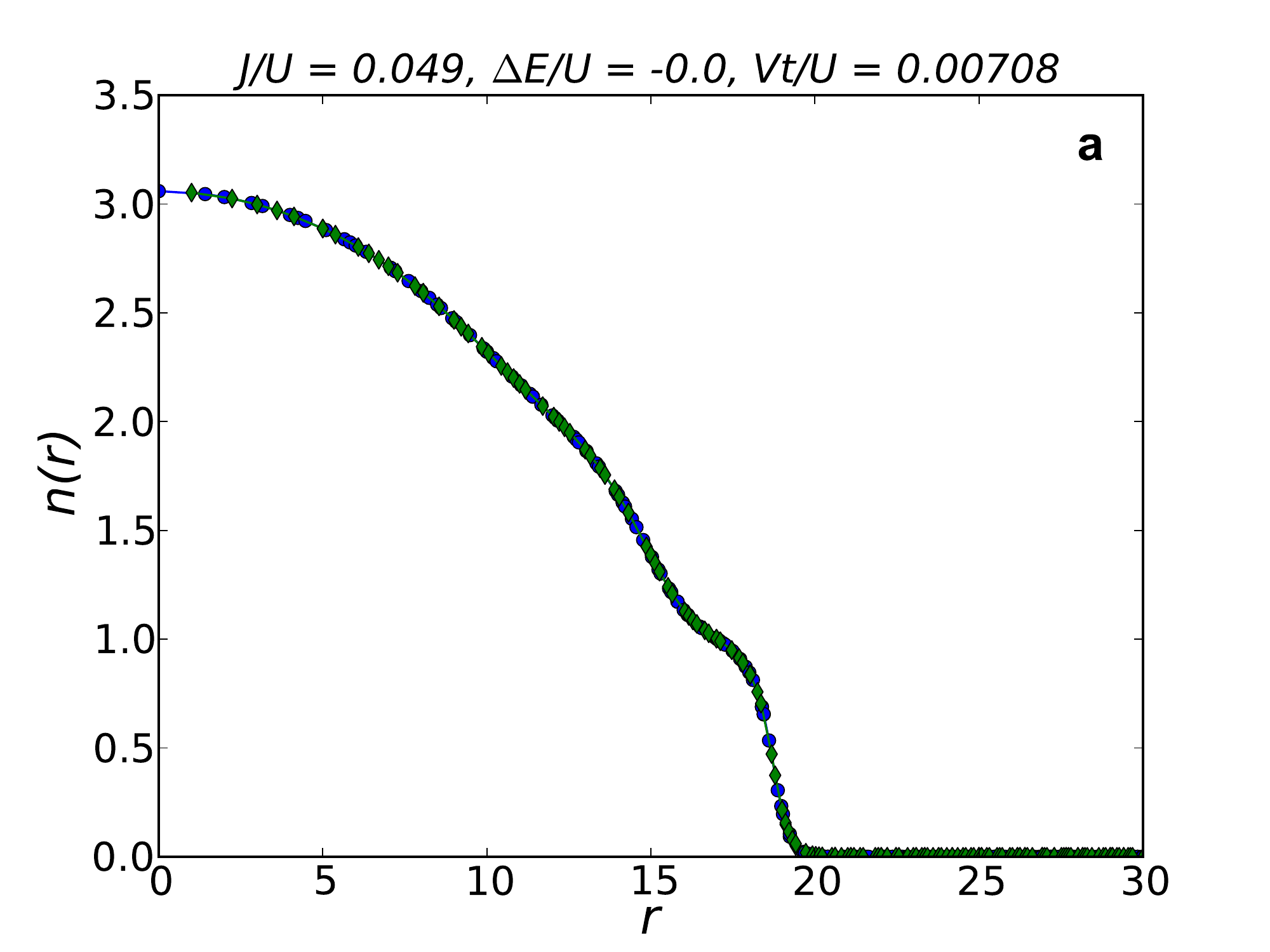}
	\includegraphics[width=0.45\textwidth]{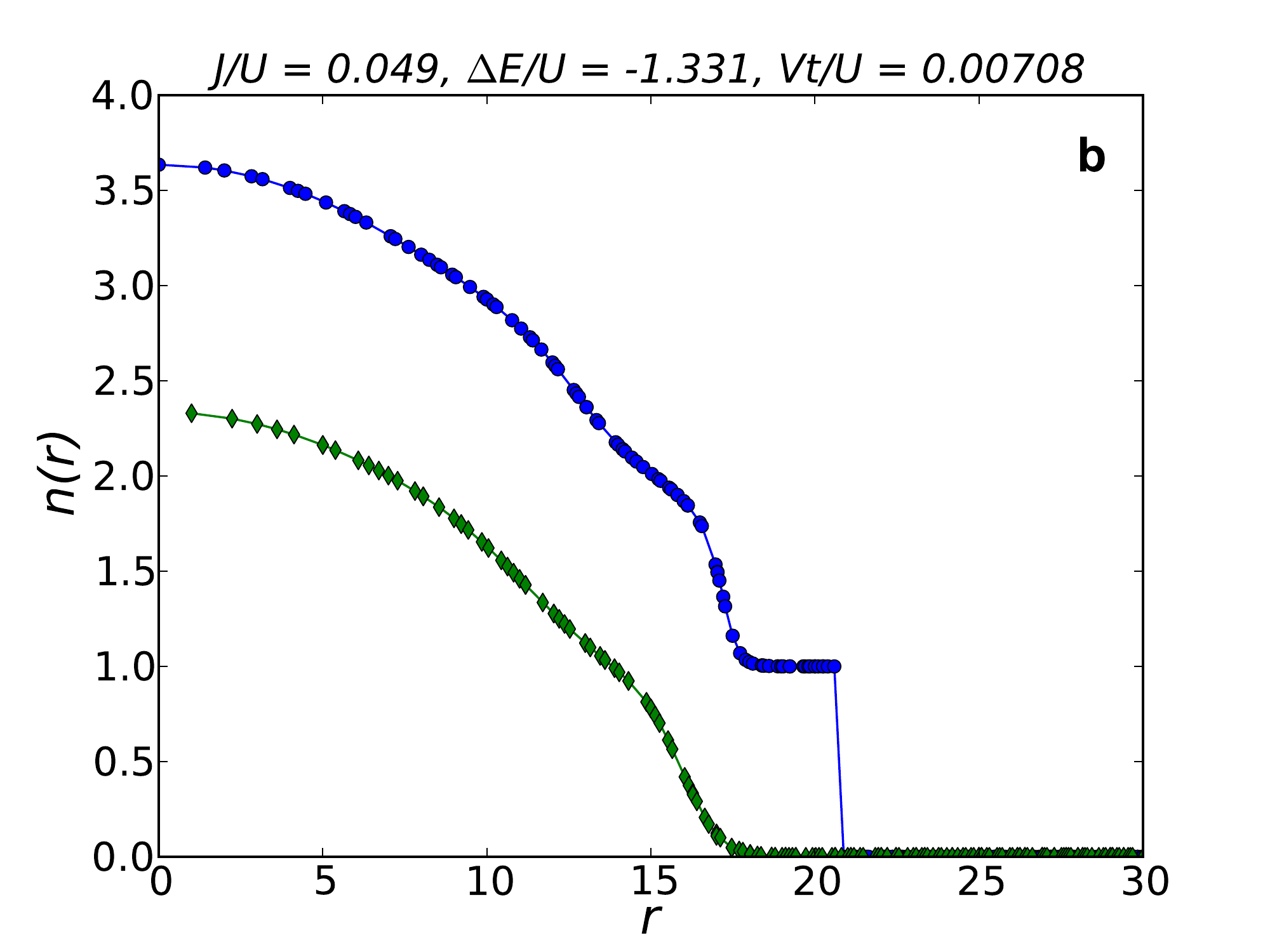}\\
	\includegraphics[width=0.45\textwidth]{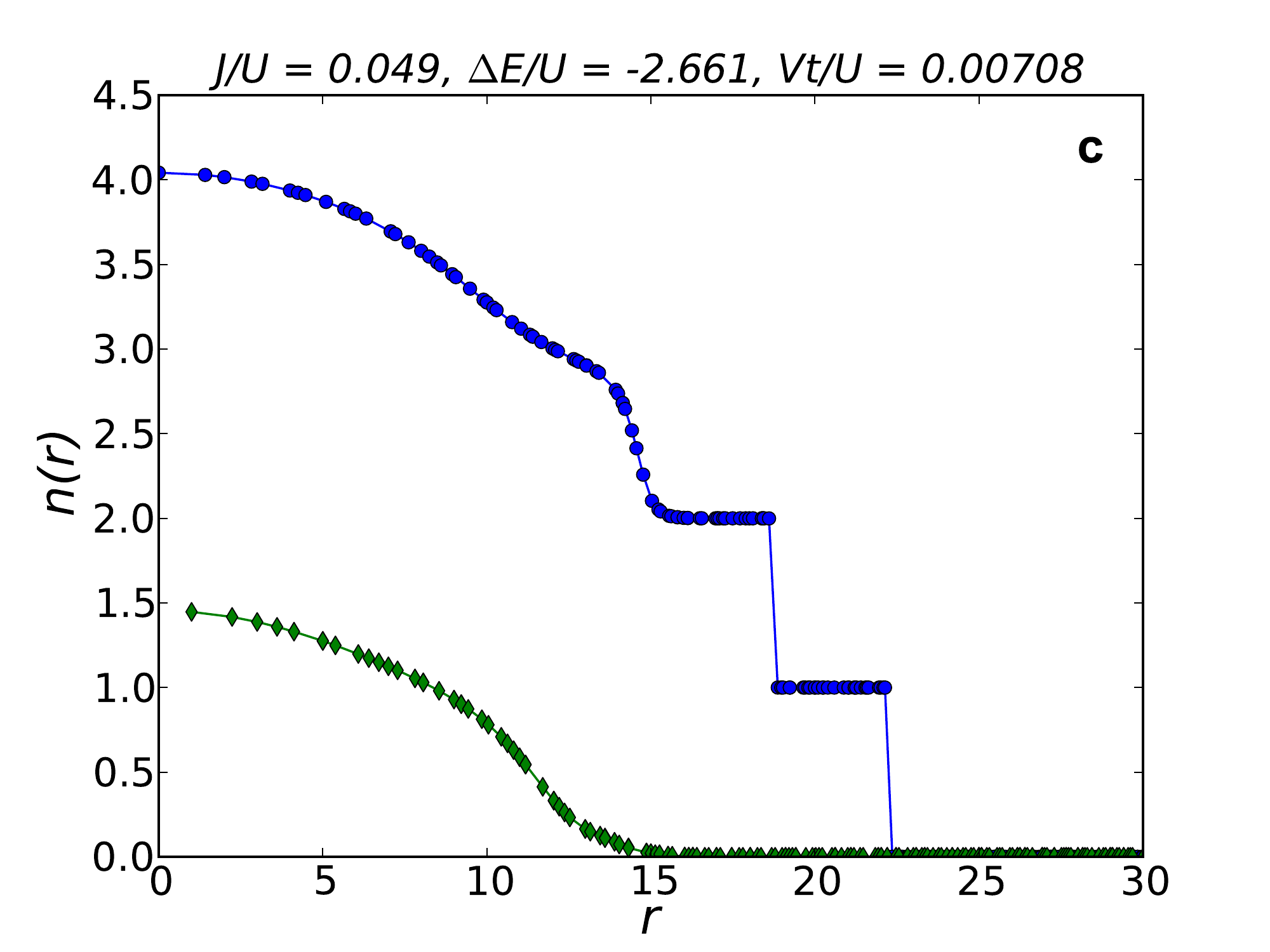}
	\includegraphics[width=0.45\textwidth]{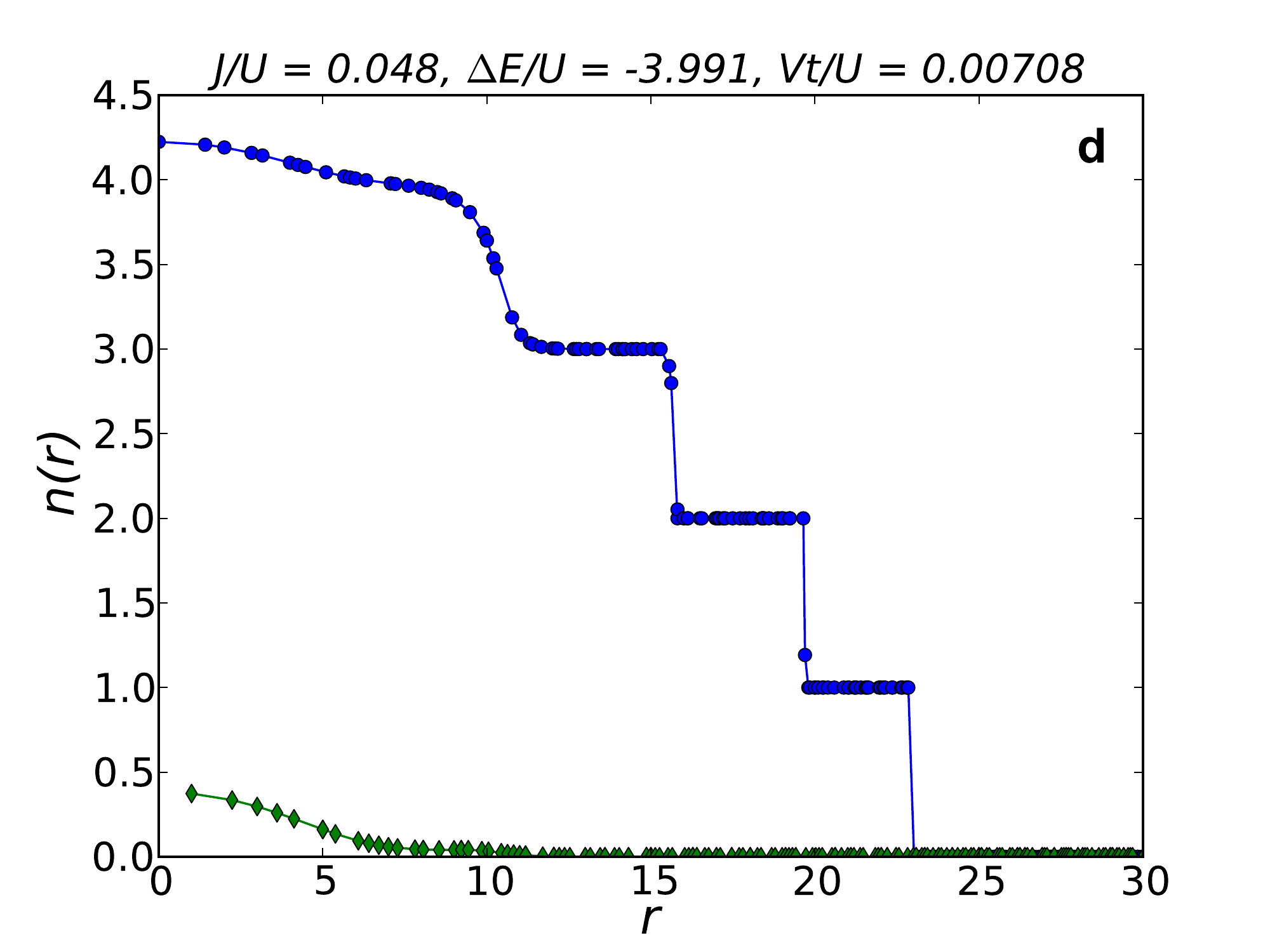}\\
	\includegraphics[width=0.45\textwidth]{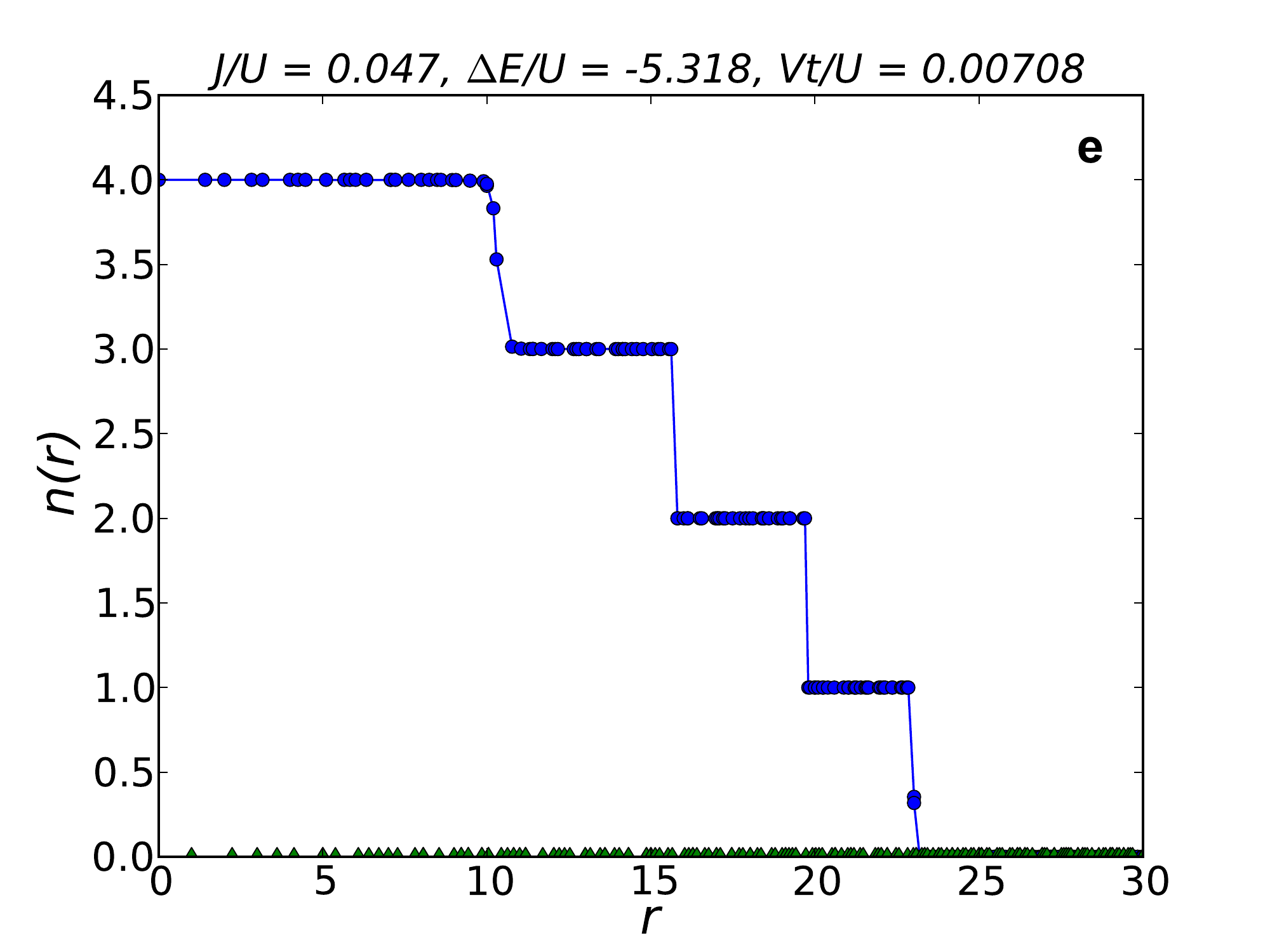}
	\includegraphics[width=0.45\textwidth]{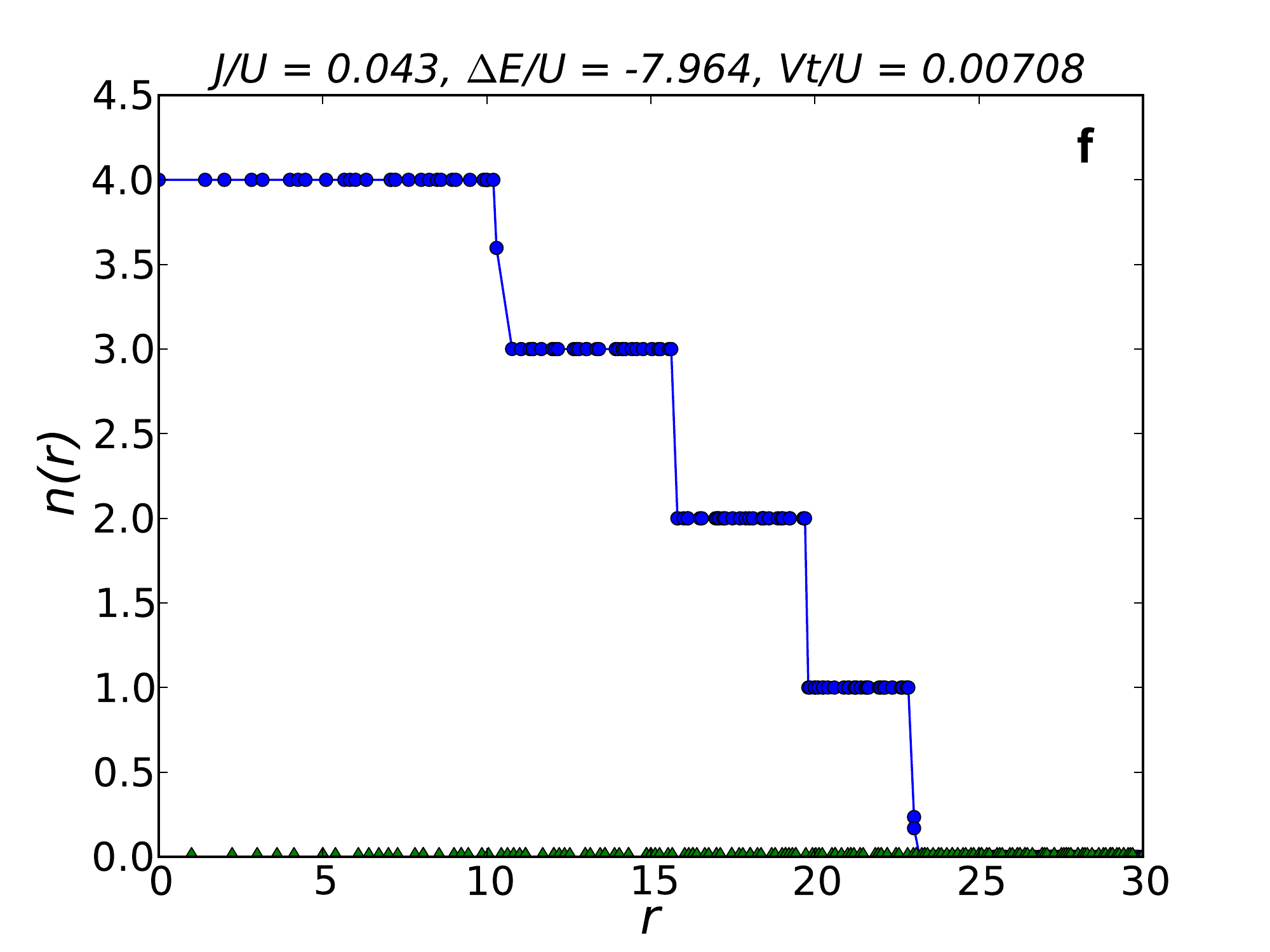}
 	\end{center}
 	\caption{Density profiles obtained using the Gutzwiller ansatz for an extended $69\times 69$ lattice in the presence of an harmonic trap for (a) $\theta = 0.5$, (b) $\theta = 0.505,$ (c) $\theta = 0.51$,(d) $\theta = 0.515$, (e) $\theta = 0.52$, (f) $\theta = 0.53$ at $V_0 = 10 E_{rec}$. Circles (diamonds) denote $\mathcal{A}$  ($\mathcal B$) sites.}
 	\label{fig:gutzdataV10}
 \end{figure}

\subsection{IV. Gutzwiller scheme}

The Gutzwiller ansatz approximation used in this work is an extension of the well-known procedure employed for the Bose-Hubbard model in conventional monopartite lattices \cite{schroll2004,zakrzewski2005}. The wave-function is assumed to be a product of single-site wave-functions $|\phi \rangle = \prod_i |\phi_i\rangle$. On each site the ansatz reads
\be
|\phi_i\rangle = \sum_{n=0}^\infty f^{(i)}_{n} | n \rangle\,,
\ee
where $i=A,B$. We have included states up to $n = 7$ and considered real Gutzwiller coefficients, which is allowed because of the U(1) symmetry and the fact that the ground state cannot have nodes, according to Feynman's no-node theorem. 

\begin{figure}[!htbp]	
 	\begin{center}
 	\includegraphics[width=0.4\textwidth]{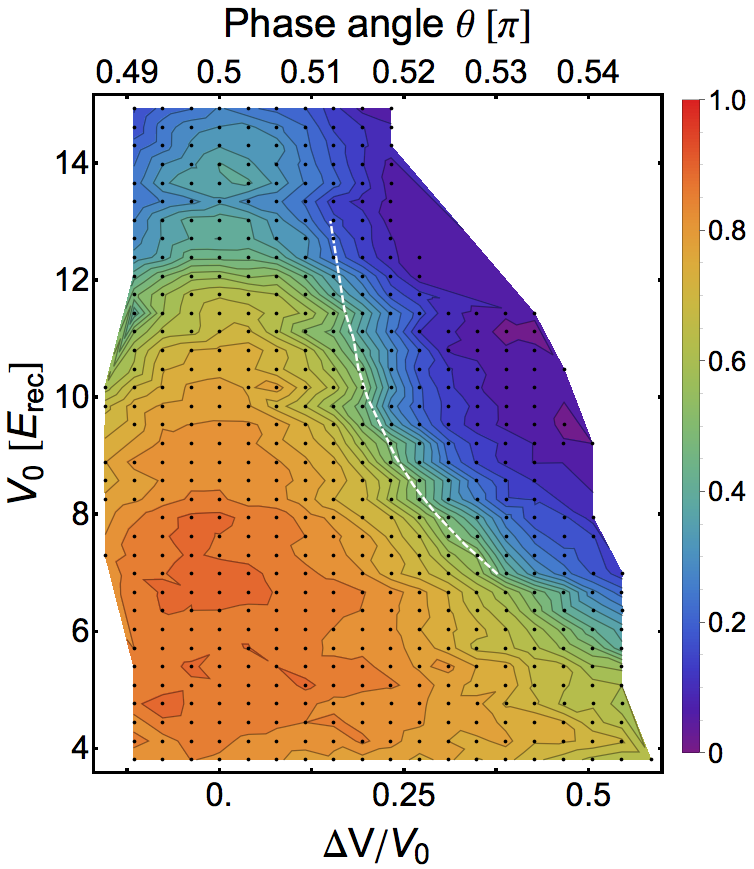}
 	\end{center}
 	\caption{Experimental visibility data. The values of $\Delta V/V_0$ and $V_0$ where measurements were taken are indicated by the black dots. The measured visibility is parametrized by the color scale on the right. The solid light-grey lines show interpolation contours.The white dashed line is a theoretical result for the critical values at which the population of the ${\cal B}$ sites vanish, obtained without fitting parameters.}
 	\label{fig:rawdata}
 \end{figure}

As shown in Sec.~II, the mean-field Hamiltonian can be written as a sum of site decoupled local Hamiltonians represented in the local Fock basis. Each local Hamiltonian needs, as an input, the order parameters of the neighbor sites ($\psi_B$ for the local Hamiltonian on site $\mathcal{A}$ and vice versa). One can thus use the following iterative procedure to determine the ground state at a given value of $J/U$ and $\mu_{A,B}/U$: one starts with a random guess of the order parameters $\psi_{A,B}$, diagonalizes the local Hamiltonians $H_A$ and $H_B$, takes the eigenvectors of the lowest energy state (i.e., the Gutzwiller coefficients $f^{(i)}_n$), calculates the new order parameters $\psi_i = \langle a^\dagger_i \rangle = \sum_n \sqrt{n+1} f^{(i)}_n f^{(i)}_{n+1}$ and repeats until convergence. In this way, we have obtained Fig.~3 in the main text and Fig.~\ref{fig:gutzdataV10} for the density profiles. By collecting the points where $n_B$ vanishes, as a function of $\Delta V/V_0$, for several values of $V_0$, we find the white line plotted in Fig.~2 of the main text. 
Here, we show a similar plot, however, in addition to interpolating contours, we also show the points where measurements have been taken (see Fig.~\ref{fig:rawdata}).

\subsection{V. Perturbative results for the visibility in the asymptotic limit.}

The regime where the imbalance between $\cal A$ and $\cal B$ sites is large can be studied using perturbation theory up to second order \cite{sakurai} when the filling is chosen to be integer in the homogeneous case. In the limit where the hopping term is neglected (which is also the mean-field ground state), the ground state is given by a perfect Mott insulator of the form $(g_A,g_B) = (g,0)$
\be
|MI\ket = \prod_{i\in \mathcal{A}} | g \ket_i \prod_{j\in \mathcal{B}} | 0\ket_j\,.
\ee
Let us start from the first order term. The only non-vanishing terms are the ones for which a particle is removed from a site $\cal A$ and moved to one of the nearest-neighbor $\cal B$ sites. The energy difference is $\Delta =  U(g-1) + \Delta \mu$ and the first order correction has thus the form
\be
-\f{J}{\Delta}\sum_{\mv{i,j}} a^\dag_i a_j |MI\ket\,.
\ee
The quadratic correction is such that a particle is removed from an $\cal A$ site, moved to a nearest-neighbor $\cal B$ site and from there it is transferred again to an $\cal A$ site which is different from the original one. The final $\cal A$ site can be a nearest-neighbor $\cal A$ site or a next-nearest-neighbor one. The correction becomes
\be
-\f{2J^2}{U\Delta}\sum_{\mv{i,j}_A} a^\dag_i a_j |MI\ket  -\f{J^2}{U\Delta}\sum_{\mv{\mv{i,j}}_A} a^\dag_i a_j |MI\ket\,.
\ee
The ground state is therefore
\be
\label{grst}
|\psi_G\ket = \left(1 - \f{J^2}{2\Delta^2} \right)|MI \ket -\f{J}{\Delta}\sum_{\mv{i,j}} a^\dag_i a_j |MI \ket -\f{2J^2}{U\Delta}\sum_{\mv{i,j}_{\cal A}} a^\dag_i a_j |MI \ket  -\f{J^2}{U\Delta}\sum_{\mv{\mv{i,j}}_{\cal A}} a^\dag_i a_j |MI \ket\,, 
\ee
where the first term is simply the unperturbed term with a wave function renormalilzation.  
\begin{figure}[!htbp]	
 	\begin{center}
 	\includegraphics[width=0.4\textwidth]{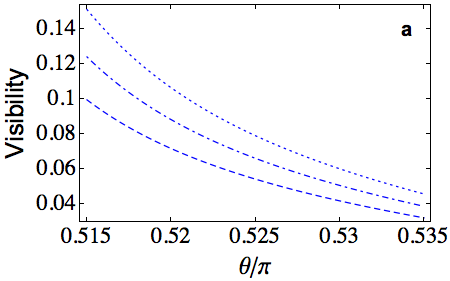}
	\includegraphics[width=0.4\textwidth]{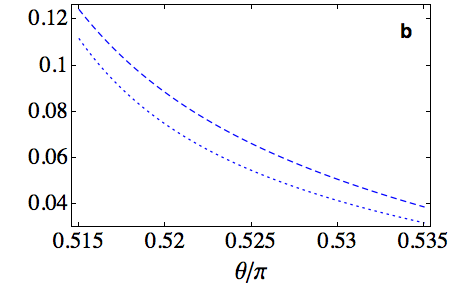}
 	\end{center}
 	\caption{Visibility curves calculated within perturbation theory for $V_0 = 10.8\, E_{rec}$. (a) Visibility up to second order for average filling $\bar g =2$ (dashed), $\bar g=2.5$ (dash-dotted), $\bar g=3$ (dotted). (b) Comparison of the visibility curves including contributions up to first order (dotted) and up to second order (dashed) for average filling $\bar g=2.5$.}
 	\label{fig:visib}
 \end{figure}
 
 \begin{figure}[!htbp]	
 	\begin{center}
 	\includegraphics[width=0.4\textwidth]{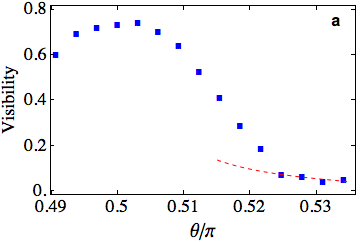}
	\includegraphics[width=0.4\textwidth]{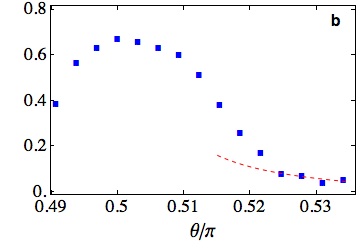}
 	\end{center}
 	\caption{Visibility data for (a) $V_0 = 10.8\,E_{rec}$ and (b) $V_0 = 11.44\,E_{rec}$. The red dashed line is obtained by fitting the last four data points with Eq.~(\ref{visib}) using the average filling $\bar g$ as a fitting parameter. We obtain (a) $\bar g=2.75\pm0.23$ and (b) $\bar g=3.77\pm0.31$.}
 	\label{fig:datavisib}
 \end{figure}
 
We can now calculate the momentum distribution
\be
S(\b k) = \f 1 N_s \sum_{i,j} e^{i\b k \cdot(\b r_i - \b r_j)}\mv{a^\dag_i a_j}\,,
\ee
where $N_s$ is the number of unit cells in the system. The visibility $\cal{V}$ is calculated at momenta $k_{max} = (0,0)$ and $k_{min} = (\sqrt{2}\pi,\sqrt{2}\pi)$. Therefore, 
\bea
S_{max} &=& \left(1 - \f{J^2}{\Delta^2} \right) g - 8 g(g+1) \f J \Delta\left( \f {3J} U + 1 \right)\,,\\
S_{min} &=& \left(1 - \f{J^2}{\Delta^2} \right) g -4 g(g+1) \left[ \f J \Delta(r_1 + 1) + \f{2J^2}{U\Delta}(2r_1 + r_2) \right]\,,
\eea
where $r_1\equiv \cos(\sqrt 2 \pi) \approx -0.266$ and $r_2\equiv \cos(\sqrt 8 \pi) \approx -0.858$, and we eventually find 
\bea
\label{visib}
\mathcal{V} &=& (S_{max} - S_{min} ) / (S_{max} + S_{min} )\nonumber\\
&=& - 2 (g+1) (1-r_1) \f J \Delta + (g+1)(2r_1 + r_2 - 3)\f{4J^2}{\Delta U } - 4(g+1)^2(r_1+3)(1-r_1)\f {J^2} {\Delta^2}\,.
\eea
The visibility obtained in Eq.~(\ref{visib}) is of the order $10^{-1}$ in the highly imbalanced regime for filling between 2 and 3 (see Fig.~\ref{fig:visib}(a)). The second order processes contribute significantly, as can be observed in Fig.~\ref{fig:visib}(b). In the theory just discussed, we did not include the contributions given by the bare next-nearest-neighbor hopping processes ($J_A$), despite the fact that the ground state (\ref{grst}) effectively includes this type of hopping contributions through virtual transitions. The reason is that the effective hopping processes contribute more substantially to the visibility than the bare ones (not displayed here).

In Fig.~\ref{fig:datavisib}, the experimental data for the visibility are plotted for $V_0 = 10.8\,E_{rec}$ and $V_0 = 11.44\,E_{rec}$. The behavior of the visibility at large imbalance, where the system is deeply in a Mott insulator phase, can be described by Eq.~(\ref{visib}), where the average filling $\bar g$ has been used as a fitting parameter. 

\subsection{VI. Quantum Monte Carlo results in 2D}

Here, we describe the QMC procedure used to obtain the results for the critical values of the interaction at the tip of the $g=1,2,3$ lobes in the phase diagram of the homogeneous Bose-Hubbard model for the monopartite lattice. We make use of the worm algorithm, as implemented in the ALPS libraries \cite{albuquerque2007,bauer2011}. By measuring the superfluid stiffness, we are able to distinguish between the two phases of the homogeneous system on a square lattice. We use a finite-size scaling to determine the position of the critical point, keeping the product of the temperature $T$ and the linear size $L$ of the lattice constant \cite{smakov2005}: $T\times L = 0.1\, U$. The comparison with a precise QMC calculation at zero temperature for filling $g=1$ \cite{capogrosso-sansone2008} allows us to conclude that the choice of temperature is adequate to describe the zero-temperature system.

\begin{figure}[htb]
\centering
\includegraphics[width=\linewidth]{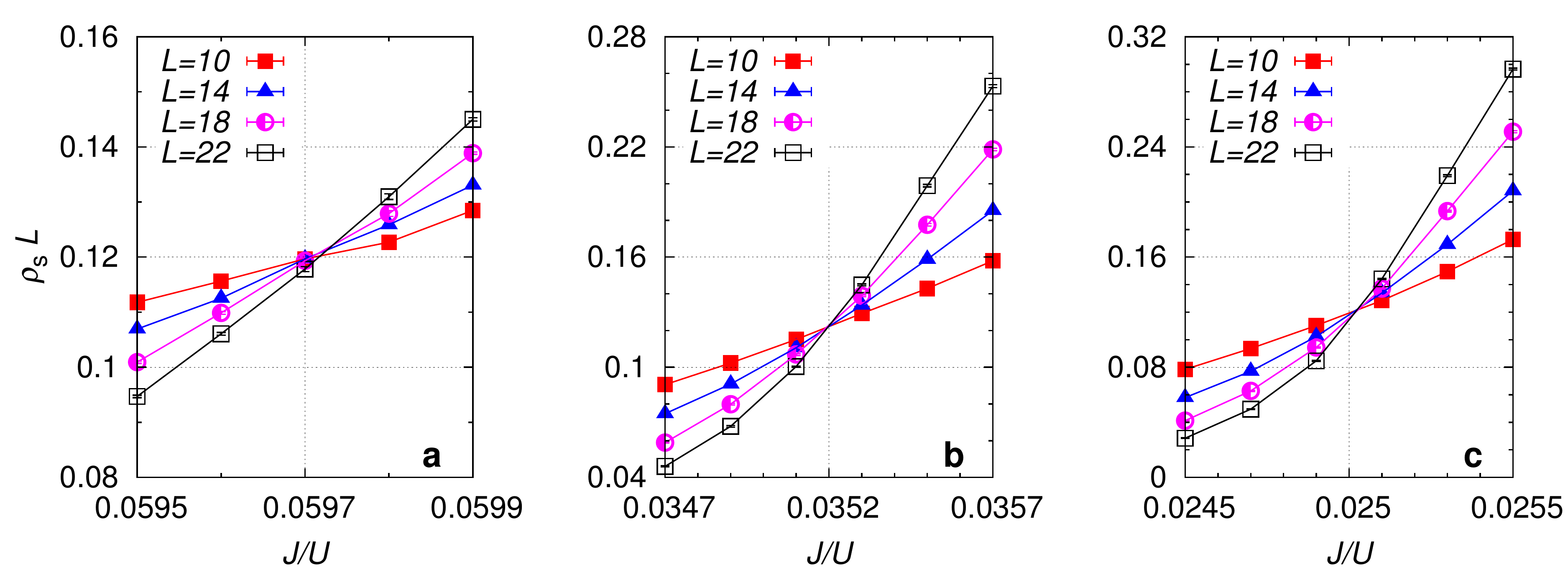}
\caption{Finite-size scaling of the superfluid stiffnes. Values of the product of the superfluid stiffness and the system size $\rho_S L$ are shown as a function of $J/U$, for different values of the linear size $L$. The three plots corresponds to the tips of the lobes with (a) $g=1$, (b) $g=2$ and (c) $g=3$. Statistical errors are smaller than symbol sizes.
\label{fig-scaling}}
\end{figure}

We study the system for different values of the ratio $J/U$, while keeping the chemical potential constant and equal to $\mu/U=0.371,1.427,2.448$, for $g=1,2,3$, respectively.
The choice of $\mu$ for $g=1$ is comparable with the results in Ref.~\cite{capogrosso-sansone2008},  while the choices of $\mu$ for  $g=2$ and $g=3$ are taken from Ref.~\cite{teichmann2009}. We set the maximum on-site occupation number to be $g+2$ (and $g+3$ for $g=1$), thus allowing for more processes than just particle-hole excitations.

Our goal is to evaluate the lobe positions using a more reliable method than the usual mean-field approach \cite{vanoosten2001}. We stress that a higher precision in the determination of the critical points, that could be obtained by lowering the temperature, increasing the system size and using a finer scan of the area around the lobe tip, is beyond the scope of this work, as it would not be relevant in the comparison with experimental results. For $g=1,2,3$, we find the following values of $(J/U)_c$: $0.0597\pm 0.0001$, $0.0352 \pm 0.0001$, $0.0250\pm 0.0001$, where the errors are due to the use of a finite grid for $J/U$. These values are in agreement with a high-precision $T=0$ result for $g=1$ \cite{capogrosso-sansone2008}, and with estimates given in Ref. \cite{teichmann2009}, based on the use of the effective potential method and Kato's perturbation theory (we observe that the latter  values of $(J/U)_c$ are systematically smaller than the ones we find). In Fig.~\ref{fig-scaling}, we show the finite-size scaling, as done in Ref.~\cite{smakov2005}.

\subsection{VII. Analogies to high-$T_c$ superconductors}

Our work may shed some light also on the behavior of similar condensed-matter systems, where loss of phase coherence occurs due to a structural modification of the lattice. 
One possible example are high-$T_c$ cuprates. 
Although the phenomenon of superconductivity occurs due to paired electrons, and here we are studying bosons, our system could bear some similarities with the cuprates, if one considers the scenario of pre-formed Cooper pairs at a higher temperature scale, as suggested by several theoretical and experimental works \cite{Yaz:08,guy:05,EK,Ran}. In this case, the onset of superconductivity at $T_c$ would correspond simply to phase coherence of the pre-formed "bosons". 

\begin{figure}[htb]
\centering
\includegraphics[width=0.3\linewidth]{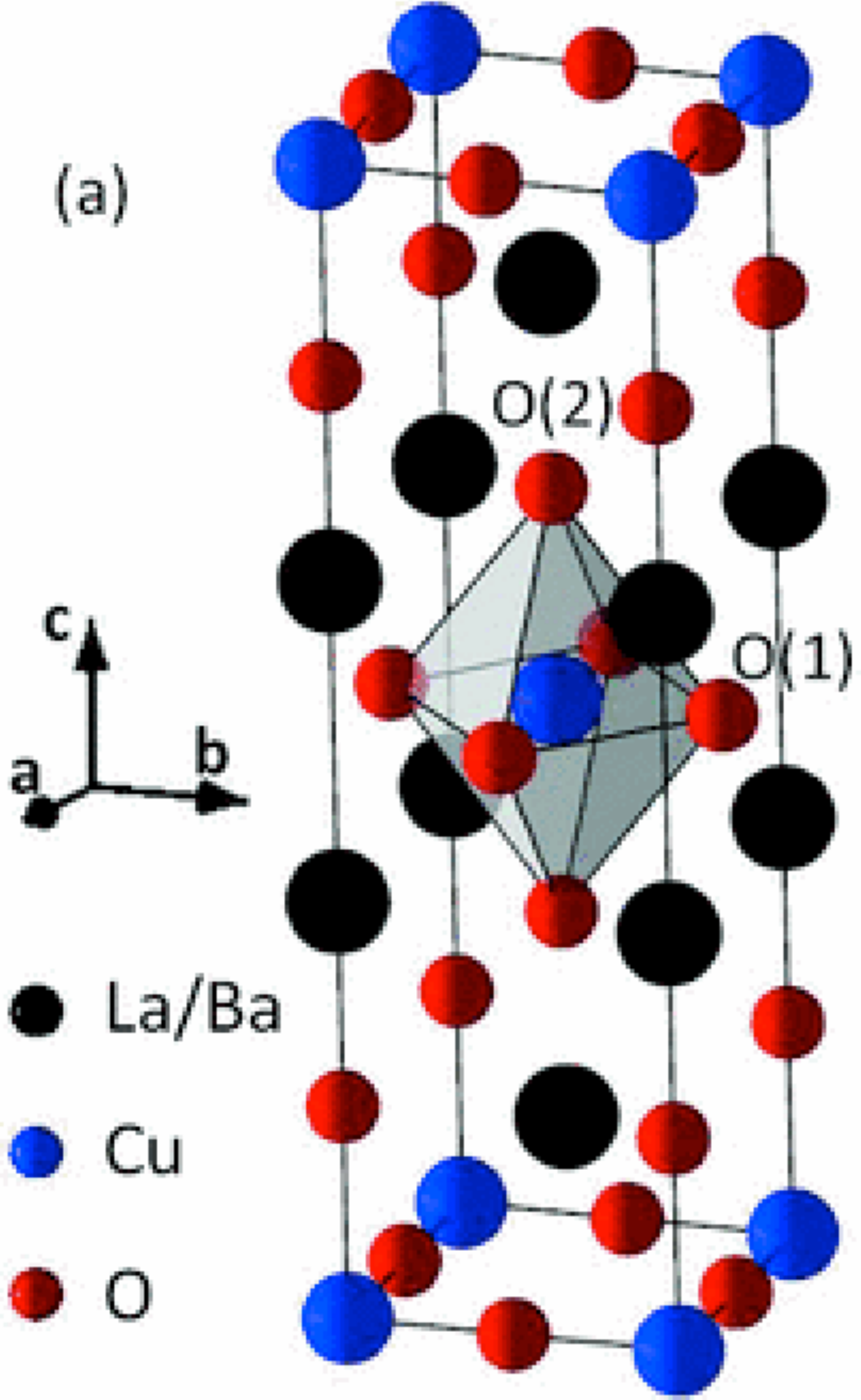}
\includegraphics[width=0.5\linewidth]{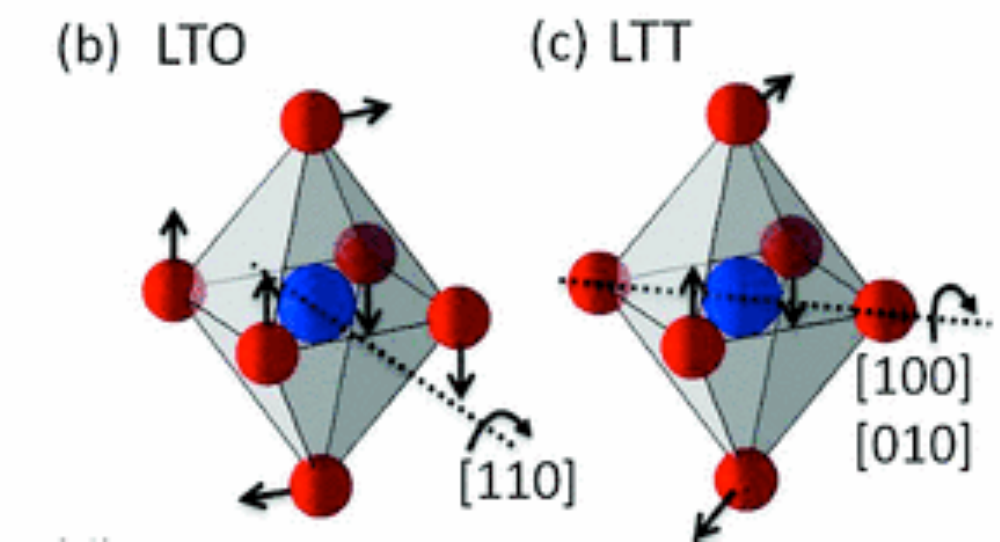}
\caption{(a) Crystal structure of  La$_{2-x}$Ba$_x$CuO$_4$ or  La$_{2-x}$Sr$_x$CuO$_4$; (b) buckling of the oxygen octahedra in the LTO and LTT phases. Figure extracted from Ref.~\cite{Haskel}
\label{crystal}}
\end{figure}

The first discovered high-$T_c$ cuprate, La$_{2-x}$Ba$_x$CuO$_4$ (see Fig.~\ref{crystal}) was found to exhibit a dip in the critical temperature at the doping value $x = 1/8$. Later, the same phenomenon was shown to occur for La$_{2-x}$Sr$_x$CuO$_4$ when La was partially substituted by some rare earth elements, like Eu or Nd \cite{Tra:13}. This feature was long known as the $1/8$ mystery, but further investigations of the materials have shown that it is connected to a structural transition from a low-temperature orthorhombic (LTO) into a low-temperature tetragonal (LTT) phase \cite{1/8}, see also Fig.~\ref{crystal}. This structural transition corresponds to a buckling of the oxygen octahedra surrounding the copper sites, which changes the nature of the copper-oxygen lattice unit cell \cite{1/8}. By increasing the concentration $y$ of Nd in La$_{2-x-y}$Nd$_y$Sr$_x$CuO$_4$, superconductivity is eventually destroyed. The onset for the disappearance of superconductivity depends also on the Sr doping $x$, but actually there is a universal critical angle $\theta_c = 3.6 \deg$ for the buckling of the oxygen octahedra, after which superconductivity cannot survive \cite{Kampf}. 

Until now, most of the theoretical studies of high-$T_c$ cuprates have concentrated on the 2D square copper lattice, but it is well known that the actual superconducting plane is composed of copper and oxygen forming  a Lieb lattice, and that the dopants sit on the oxygen (see Fig.~\ref{crystal}). 
The role of the LTO/LTT structural transition is mostly to shift two of the four in-plane oxygen atoms, which were slightly out of the plane, back into it. Although essentially more complicated than the problem studied here, the critical buckling angle $\theta_c = 3.6 \deg$ for the destruction of superconductivity \cite{Kampf} bears similarities with the critical deformation angle $\theta_c$ (or equivalently $\Delta V_c$) that we found in this work. We hope that our results will foster further investigations of the specific role played by the oxygen lattice in high-$T_c$ superconductors, and its importance in preserving phase coherence. 

\end{document}